\documentclass{aa}
\bibpunct{(}{)}{;}{a}{}{,}
\usepackage[varg]{txfonts}
\usepackage{natbib,twoopt}
\usepackage{graphicx}   % Including figure files
\usepackage{amsmath}    % Advanced maths commands
\usepackage{amssymb}    % Extra maths symbols
\usepackage[usenames,dvipsnames]{xcolor}
\usepackage{multirow}
\usepackage{amsfonts}
\usepackage{mathrsfs}
\usepackage{nccmath}
\usepackage{verbatim}
\usepackage{lscape}
\usepackage{ulem}
\usepackage{float}
\usepackage{placeins}
\usepackage{stfloats}
\usepackage{booktabs}
\usepackage{longtable}
\usepackage{newtxtext,newtxmath}%%%%%%%%%%%%%%%%%%%%%%%%%%%%%%%%%%%%%%%%%%%%%%%%%%
\usepackage{float}
\usepackage{soul}

\newcommand{\Msun}[0]{\rm{M_\odot}}
\newcommand{\FeH}[0]{\rm{[Fe/H]}}
\newcommand{\MH}[0]{\rm{[M/H]}}

\newcommand{\Teff}[0]{T_\mathrm{eff}}

\begin{document}

\title{Characterisation of FG-type stars with an improved transport of chemical elements}
%\subtitle{}
\author{Nuno Moedas\inst{1,2}\thanks{nmoedas@astro.up.pt} \and Diego Bossini\inst{1} \and Morgan Deal\inst{3} \and Margarida Cunha\inst{1}}

\institute{ Instituto de Astrof\'isica e Ci\^encias do Espaço, Universidade do Porto, CAUP, Rua das Estrelas, PT4150-762 Porto, Portugal
\and Departamento de F\'isica e Astronomia, Faculdade de Ci\^encias da Universidade do Porto, Rua do Campo Alegre, s/n, PT4169-007 Porto, Portugal, \and LUPM, Universit\'e de Montpellier, CNRS, Place Eug\`ene Bataillon, 34095 Montpellier, France}

\date{Received XXX / Accepted YYY}

\abstract {The modelling of chemical transport mechanisms is crucial for accurate stellar characterizations.  Atomic diffusion is one of these processes and it is commonly included in stellar models. 
%This mechanism is known to be important for the modelling of solar-type stars.
However, it is usually neglected for F-type or more massive stars because it produces surface abundance variations that are unrealistic. Additional mechanisms to counteract atomic diffusion must therefore be considered. It has been demonstrated that turbulent mixing can prevent the surface abundance over-variations, and can also be calibrated to mimic the effects of radiative accelerations on iron.}{We aim to evaluate the effect of a calibrated turbulent mixing on the characterisation of a sample of F-type stars, and how the estimates compare with those obtained when the chemical transport mechanisms are neglected.}{We selected stars from two samples - one from the \textit{Kepler} LEGACY sample and the other from a sample of \textit{Kepler} planet-hosting stars. We inferred their stellar properties using two grids. The first grid considers atomic diffusion only in models that do not show chemical over-variations at the stellar surface. The second grid includes atomic diffusion in all the stellar models and the calibrated turbulent mixing to avoid unrealistic surface abundances.}{Comparing the derived results from the two grids, we found that the results for the more massive stars in our sample will have higher dispersion in the inferred values of mass, radius and age, due to the absence of atomic diffusion in one of the grids. This can lead to relative uncertainties for individual stars of up to 5\% for masses, 2\% for radii and 20\% for ages.}{This work shows that a proper modelling of the microscopic transport processes is key for an accurate estimation of their fundamental properties not only for G-type stars, but also for F-type stars. } 
\keywords{Diffusion - Turbulence - Stars: abundances - Stars: evolution - Asteroseismology}

\titlerunning{Transport of chemical elements}
\authorrunning{Nuno Moedas \& et al.}
\maketitle

\section{Introduction}
\label{sec:Intro}

%\dbc{The tabled in the appendix need to be cited somewhere, even in the main text. Plus, fix the position of the tables and the appendix titles: }

An accurate and precise characterisation of stars is fundamental to better understand the evolution of the Universe. Advances were made thanks to the high-quality data provided by missions like CoRoT/CNES \citep{corotteam2006}, \textit{Kepler}/K2 \citep{kepler} and the Transiting Exoplanet Survey Satellite (TESS; \citealt{TESS}). 
These missions provided new constraints with asteroseismology linked to the stellar interior and evolution. Missions that will be launched in the near future, such as PLAnetary Transits and Oscillations of stars (PLATO/ESA; \citealt{Rauer2014}), will enrich even further the knowledge about stars with the precise data that they will provide. 

However, it is currently difficult to achieve the accuracy requirements imposed by the future mission PLATO in terms of masses, radii and particularly ages (it requires 10\% accuracy in age for stars similar to the Sun; \citealt{Rauer2014}).
This is due to our lack of knowledge and approximations made in the modelling of the physical processes taking place inside stellar models.
One source of uncertainties is the modelling of chemical transport mechanisms acting inside stars. These processes can be either microscopic or macroscopic and can compete with each other, leading to a redistribution of the chemical elements inside a star. This affects the internal structure, evolution and abundance profiles of stars. 
Atomic diffusion is one of these processes. This microscopic transport process is driven mainly by pressure, temperature and chemical gradients, redistributing the elements throughout the stellar interior \citep{michaud15}.  \cite{Valle2014,Valle2015} tested the impact of diffusion on the stellar properties and found that neglecting it can lead to uncertanties of 4.5\%, 2.2\% and 20\% for mass, radius and age.
In a model-based controlled study performed  in the context of PLATO, \cite{Cunha2021} found that atomic diffusion can impact the inferred age accuracy on the order of 10\%,  for a 1.0~M$_\odot$ star close to the end of the main sequence. Furthermore, using real data from Kepler, \cite{Nsamba2018} found a systematic difference of 16\% in the age inferred from grids with and without diffusion on a sample of stars with masses smaller than $M=1.2$~M$_\odot$. These results emphasize the need to enhance our understanding and modelling of atomic diffusion and other chemical transport mechanisms.

Atomic diffusion can be decomposed into two main competing sub-processes. One is gravitational settling, which brings the elements from the stellar surface into the deep interior, except for hydrogen which is transported from the interior to the surface. The other is the radiative acceleration that pushes some elements, mainly the heavy ones, towards the surface of stars due to a transfer of momentum between photons and ions. 
Several studies showed that the efficiency of the processes depends on the stellar fundamental properties which translate into an increase of the efficiency with mass and a decrease with metallicity \citep[see e.g.][and references therein]{Deal2018,moedas2022}. 
 Although works of \cite{Chaboyer2001,Salaris2001} prove that gravitational settling is successful alone to predict the surface abundances of low-mass stars, the effects of radiative accelerations become important for stars with a small surface convective zone (e.g. for solar-metallicity stars with an effective temperature higher than $\sim6000$~K, \citealt{michaud15}). Nevertheless, atomic diffusion alone for stars more massive than the Sun causes variations on the surface abundances that are larger than the ones observed in clusters \citep[e.g.][]{gruyters14,Gruyters2016,Semenova2020}. This indicates the need of additional chemical transport mechanisms, like the radiative accelerations.  However, radiative acceleration is highly computationally demanding and therefore is often neglected in stellar models \citep{GARSTEC,PARSEC,BASTI1,BASTI2}.

Some works \citep[e.g.][and references therein]{Eggenberger2010,Vick2010,Deal2020,dumont20} proved the necessity of including other chemical transport processes in competition with atomic diffusion. Nevertheless, the identification and accurate modelling of the different processes are still ongoing. 
The processes that  can be considered are either diffusive or advective. If we assume that all of them are fully diffusive we can parameterise their effects by considering a turbulent mixing coefficient, that can be constrained using the surface abundances of stars in cluster \citep{Gruyters2013,Gruyters2016,Semenova2020}. This has also been performed in F-type stars by \cite{Verma2019}, where they used the glitch induced by the helium second ionisation region to calibrate the turbulent mixing coefficient that best reproduced the helium surface abundances. \cite{Eggenberger2022} also showed that the effect of the rotation-induced mixing in the Sun could be parameterised with a simple turbulent diffusion coefficient expression.
More recently, \cite{moedas2022} found that it is possible to add the effects of radiative accelerations on iron into the turbulent mixing calibration. That work showed that this calibration depends on the stellar mass (as the mass increases the value of turbulent mixing is increased to mimic the effect of radiative accelerations). Such parameterisation of the transport should improve the determination of stellar mass, radius and especially age. It allows to include atomic diffusion avoiding the unrealistic surface abundance variation (for non-chemically peculiar stars) it would induce alone. However, it was also showed that, as expected, the turbulent mixing calibration is not able to reproduce the evolution of all chemical elements. Nevertheless, it reproduces the abundance of iron which is the main element used as observational constraint in stellar models, allowing a global characterisation of stars. Of course the calibration is only valid for a given physics and it should be redone when it is changed or when the initial chemical composition is different, especially for different alpha-enhancement.

In this work, we use the calibration presented in \cite{moedas2022} to characterise a sample of FG seismic stars, selected from the \textit{Kepler} LEGACY sample \citep{Lund2017} and the planet-host stars studied by \cite{Davies2016}. We use these stars to see how the calibrated turbulent mixing performs in stellar characterisation and to see how the results compare to the determinations obtained with standard models for F-type stars (i.e. without atomic diffusion).

This article is structured as follows. In Sect. \ref{sec:models} we present the input physics of the grids of stellar models.  In Sect. \ref{sec:sample} we present the stellar sample we use. In Sect. \ref{sec:opt} we explain the optimisation process considered.  The main results of the stellar characterisation are presented in Sect. \ref{sec:results}. We discuss the results of using different seismic frequency weights, data quality, and comparison with the results of previous works in Sect. \ref{sec:discussion}. We conclude in Sect. \ref{sec:conclusion}.

\begin{table*}[ht!]
\centering
\caption{Parameter space and transport processes of the grids of stellar models. The atomic diffusion in the stellar models does not consider radiative accelerations.}
\label{tab:stellar_grids}
\resizebox{0.8\textwidth}{!}{%
\begin{tabular}{ccccccccc}
\hline
\multirow{2}{*}{Grid} & \multicolumn{2}{c}{Mass (M$_\odot$)} & \multicolumn{2}{c}{$\rm [M/H]_i$} & \multicolumn{2}{c}{$Y_i$} & \multirow{2}{*}{Atomic Diffusion}  & \multirow{2}{*}{Turbulent Mixing} \\ \cline{2-7}
 & Range & Step & Range & Step & Range & Step & &  \\ \hline
\multirow{2}{*}{A} & \multirow{4}{*}{[0.7;1.75]} & \multirow{4}{*}{0.05} & \multirow{4}{*}{[-0.4;0.5]} & \multirow{4}{*}{0.05} & \multirow{4}{*}{[0.24;0.34]} & \multirow{4}{*}{0.01} & \multirow{2}{*}{$\rm \Delta [Fe/H]_{MAX}<0.2$ } & \multirow{2}{*}{No} \\
 &  &  &  &  &  &  &  &  \\ \cline{1-1} \cline{8-9} 
\multirow{2}{*}{B} &  &  &  &  &  &  & \multirow{2}{*}{All models} & \multirow{2}{*}{$\rm D_{T,Fe}$} \\
 &  &  &  &  &  &  &  &  \\ \hline
\end{tabular}%
}
\end{table*}
\section{Stellar Models}
\label{sec:models}

\subsection{Stellar Physics}

In order to assess the impact of turbulent mixing on the stellar properties, we built two grids of stellar models. The models are computed with MESA evolutionary code (Modules for Experiments in Stellar Astrophysics, MESA r12778, \citealt{Paxton2011,Paxton2013,Paxton2015,Paxton2018,Paxton2019}) and the input physics is the same as grid D1  of \cite{moedas2022}.
We adopted the solar heavy elements mixture given by \citet{Asplund2009}, and the OPAL\footnote{\url{https://opalopacity.llnl.gov/}} opacity tables \citep{Iglesias1996} for the higher temperature regime, and the tables provided by \cite{Ferguson2005} for lower temperatures.  All table are computed  for a given mixture of metals. We use the OPAL2005 equation of state \citep{Rogers2002}. We used the \cite{Krishna1966} atmosphere, for the boundary condition at the stellar surface. We follow the \cite{Cox1968} for convection, imposing the mixing length parameter $\boldsymbol{\alpha_\mathrm{MLT}=1.711}$, in agreement with the solar calibration we performed on-the-fly for both grids. In the presence of a convective core, we implemented core overshoot following an exponential decay with a diffusion coefficient, as presented in \citet{Herwig2000},  %$\alpha_\mathrm{MLT}=1.710653$
\begin{equation}
    D_{\rm ov}=D_0\exp\left(-\frac{2z}{fH_p}\right),
\end{equation}
where $D_0$ is the diffusion coefficient at the border of the convectively unstable region,  $z$ is the distance from the boundary of the convective region, $H_p$ is the pressure scale height, and $f$ is the overshoot parameter set to $f=0.01$.

In one of the grids we included turbulent mixing, using the prescription of \cite{Richer2000},
\begin{equation}
    D_\mathrm{T}=\omega D(\rm{He})_0\left(\frac{\rho_0}{\rho}\right)^n,
    \label{eq:Dturb}
\end{equation}
where $\omega$ and $n$ are constants, $\rho$ and $D(\rm{He})$ the local density and diffusion coefficient of helium, the indices 0 indicates that the value is taken at a reference depth. The $D(\rm{He})$  was computed following the analytical expression given by \cite{Richer2000}:
\begin{equation}
    D(\mathrm{He})=\frac{3.3\times10^{-15}T^{2.5}}{4\rho\ln{(1+1.{1}25\times10^{-16}T^3/\rho)}},
    \label{eq:DHe}
\end{equation}
where $T$ is the local temperature. In this work, we set $\omega$ and $n$ to $10^4$ and $4$, respectively \citep{Michaud2011a,Michaud2011b}. We considered the turbulent mixing parameterisation done by \cite{moedas2022}, where they added the effects of radiative acceleration on iron.
 They used a reference envelope mass ($\Delta M_0$, as the reference depth) to indicate how deep turbulent mixing reaches inside the star. They suggest that this parameter varies with the mass of the star as  
\begin{equation}\label{eq:param}
    \Delta M_0 \left(\frac{M^\ast}{M_\odot}\right)= 3.1\times 10^{-4} \times \left(\frac{M^\ast}{M_\odot}\right) + 2.7\times 10^{-4}.
\end{equation}
Higher values of $\Delta M_0$ result in more efficient mixing due to turbulent mixing to account for the radiative acceleration process. 
For more information see the work \cite{moedas2022} (and references therein).

\begin{figure}
    \centering
    \includegraphics[width=0.97\columnwidth]{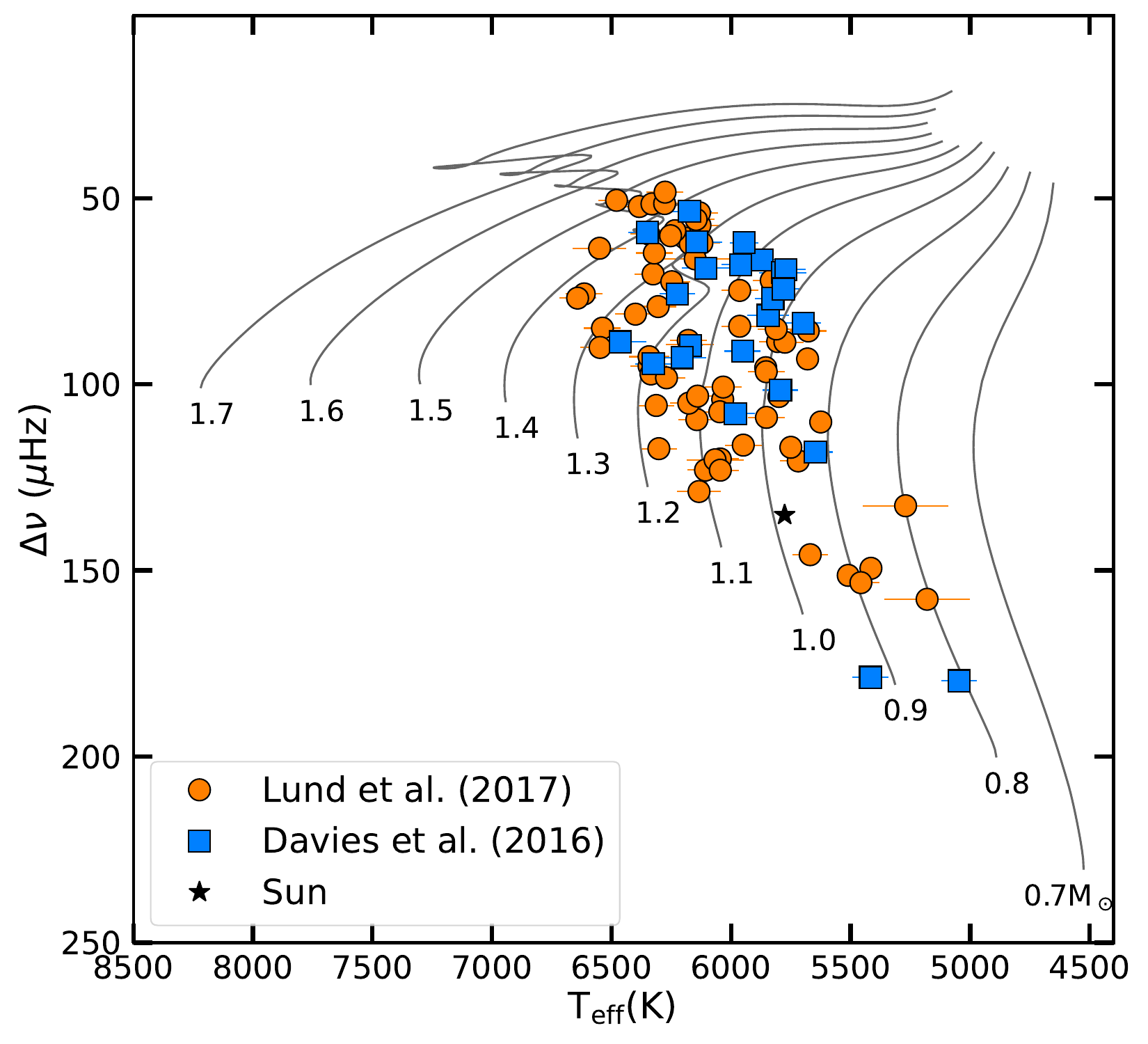}
    \caption{The Asteroseismic Diagram shows some computed evolutionary tracks with $\MH_i=0.0$ and $Y_i=0.26$ (that are not the Solar values) in solid black lines. The points show the distribution of the sample considered in this work, where the orange circles are taken  from \cite{Lund2017}, the blue squares are taken from \cite{Davies2016}, and the black star is the Sun.}
    %\caption{The Kiel Diagram shows some computed evolutionary tracks with solid black lines. The points show the distribution of all samples where the orange circles are taken  from \cite{Lund2017}, the blue squares are taken from \cite{Davies2016}, and the black star is the Sun.}
    \label{fig:sample}
\end{figure}

\subsection{Parameter Space}

Both grids cover the same parameter space, ranging masses between 0.7 and 1.75 $\Msun$ in steps of 0.05 $\Msun$, initial metallicities $\MH_i${\footnote{$\MH=\log({Z}/{X})-\log({{Z}/{X}})_\odot$} from -0.4 to 0.5 dex in steps of 0.05 dex, and initial helium mass fraction $Y_i$ between 0.24 and 0.34 in steps of 0.01. The differences between the two grids are the chemical transport mechanisms.
In grid A, turbulent mixing is not included and only atomic diffusion without radiative acceleration is taken into account, only in models where maximum variation of the iron content at the surface during all the evolution is $\Delta\FeH$\footnote{$\FeH=\log({X(Fe)}/{X})-\log({{X(Fe)}/{X}})_\odot$}$>0.2$~dex. This is to avoid unrealistic over-variations caused by atomic diffusion (see \citealt{moedas2022} for more details). Also by considering $\Delta\FeH$, we take into account the effects of changing the initial chemical composition in the efficiency of atomic diffusion (the size of the convective envelope changes with it). It is therefore expected that the majority of models of low-mass stars will include atomic diffusion in this grid. All stellar models with mass lower than 1.0~$\Msun$  are not affected by this criteria and include atomic diffusion. Stars with masses higher than 1.4~$\Msun$ are all affected and atomic diffusion is not included. In between it depends on the initial chemical composition with less models including atomic diffusion as the stellar mass increases. For example, for models with a mass of 1.3~$\Msun$ only models with $\MH_i=0.5$ and $Y_i=0.24$ include atomic diffusion, and for models with a mass of 1.0~$\Msun$ only models with $\MH_i=-0.4$ and $Y_i=0.34$ do not include atomic diffusion.

%This way we expect that the majority of the models of low-mass stars include atomic diffusion in this grid, contrarily to grids that do not include atomic diffusion at a given mass, without taking into account the effect of the initial chemical composition.

In grid B, we include turbulent mixing, using the calibration done in \cite{moedas2022} where the efficiency of the  turbulent mixing increases with the stellar mass following Eq.~\ref{eq:param}. The inclusion of this mechanism allows us to avoid the effects of over-variations in the chemical abundances and to include atomic diffusion in all the stellar models of the grid.

For both grids, we saved the models from the Zero-Age Main Sequence (ZAMS) to the bottom of the Red-Giant Branch (RGB) stage. We also computed the individual frequencies for each stellar model using the GYRE oscillation code \citep{Townsend2013}. The main properties of the grids are summarized in Table~\ref{tab:stellar_grids}.

\section{Stellar Sample}
\label{sec:sample}
The sample is selected from 2 different sources. The first source is the \textit{Kepler} LEGACY sample from \cite{Lund2017} (hereafter L17) which is a sample of 66 stars with the highest signal-to-noise ratio in the seismic frequencies observed by \textit{Kepler}. The second source is the work of \cite{Davies2016} (hereafter D16) a study of 35 stars (32 of them different from L17) that are planet-hosts. For both samples, we only selected stars with $\FeH>-0.4$ to stay within the parameter space of the grids. The smaller value of $\FeH$ is -0.37~dex, note that the value of $\FeH$ is usually smaller than $\MH$ in the stellar models, reaching a difference of up to 0.04~dex, as reported in \citealt{moedas2022}, allowing the stars selected to be well within the grid parameter space. We excluded two stars from D16 that present mix-modes in the detection (KIC7199397 and KIC8684730). Finally, we included the degraded Sun as it is presented in \cite{Lund2017} as a control star. We end up with 91 stars (62 from L17, 28 from D16 and the Sun). The distribution of the full sample is presented in the Asteroseismic Diagram of Fig. \ref{fig:sample}.
 
For all the stars, we use the effective temperature ($\Teff$), iron content ($\FeH$), frequency of maximum power ($\nu_{\rm max}$), and individual seismic frequencies ($\nu_{i}$) as constraints. We use the constraints given in the respective papers, except for 13 stars of L17, for which we use the $\Teff$ and $\FeH$ values updated in \cite{morel2021}.

The method used to characterise both L17 and D16 samples is the same. Yet the quality of the seismic data is better in the L17 sample because the stars have at least 12 months more of observations by the \textit{Kepler} mission. This allows us to see how the data quality will impact the uncertainties in the fundamental stellar property inferences. This aspect is discussed in Sec. \ref{sec:unc_LEG_DAV}.

\section{Optimisation process}
\label{sec:opt}
The fundamental properties of the sample defined in the previous section are inferred using both grids in combination with the Asteroseismic Inference on a Massive Scale (AIMS, \cite{Rendle2019}) code. AIMS is an optimisation tool that uses Bayesian statistics and Markov Chain Monte Carlo (MCMC) to explore the grid parameter space  and find the model that best fits the observational constraints.
In the present work, we use the two-term  surface corrections proposed by \cite{Ball2014}, in order to compensate for the difference between theoretical and observed frequencies, due to the incomplete  modelling of the surface layers of stars.
We explored also the different ways of considering the $\chi^2$ in the optimisation. AIMS differences the contribution of the global constraints $X_i$, (in our case $\Teff$, $\FeH$ and $\nu_{\rm max}$)
\begin{equation}
    \label{eq:Chi_class} 
    \chi^2_\mathrm{global}=\sum^3_i\left(\frac{X_i^\mathrm{(obs)}-X_i^\mathrm{(mod)}}{\sigma(X_i)}\right)
\end{equation}
and the constraints from individual frequencies $\nu_{i}$
\begin{equation}
    \label{eq:Chi_seis} 
    \chi^2_\mathrm{freq}=\sum^N_i\left(\frac{\nu_{i}^\mathrm{(obs)}-\nu_{i}^\mathrm{(mod)}}{\sigma(\nu_{i})}\right).
\end{equation}
where (obs) corresponds to the observed values and (mod) corresponds to the model values.
The weight that AIMS gives to the seismic contribution can be absolute (3:N), where each individual frequency has the same weight as each global constraint
\begin{equation}
    \label{eq:Chi_tot_abs} 
    \chi^2_\mathrm{total}=\chi^2_\mathrm{freq}+\chi^2_\mathrm{global},
\end{equation}
or relative (3:3), where all the frequencies have the same weight as all the global constraints
\begin{equation}
    \label{eq:Chi_tot_rel} 
    \chi^2_\mathrm{total}=\left(\frac{N_\mathrm{global}}{N_\mathrm{freq}}\right)\chi^2_\mathrm{freq}+\chi^2_\mathrm{global},
\end{equation}
where $N_\mathrm{global}$ and $N_\mathrm{freq}$ are the numbers of global and frequencies constraints, respectively.

\cite{Cunha2021} assessed the impact of using these two different ways to consider the weight in the frequencies. The weights 3:3  synthetically inflate the uncertainties of the individual frequencies, which allows the optimisation procedure to explore more of the parameter space. However, this is not statistically correct as explained in \cite{Cunha2021}. If we want the optimisation to be statistically correct we should instead consider 3:N weights, which use the full potential of the seismic frequencies, leading to results with smaller uncertainties.  However, as discussed in \cite{Cunha2021}, the use of 3:N weights is more sensitive to an incomplete or wrong modelling of stars and can lead to results that are incompatible with the global constrains. Given the discussion in the literature around the application of weights, in this work we decided to assess the effect of using both weight options in the results (see Sec. \ref{sec:abs_vs_rel}). The properties of masses, radii and ages inferred using Grid B are provided in tables \ref{tab:abs_res_3N} and \ref{tab:abs_res_33} for both (3:N) and (3:3) frequency weights.

\section{Results}
\label{sec:results}
\subsection{The Sun}

\begin{table}
\centering
\caption{Solar properties from the optimisation process. Age of the Sun used as reference 4.57~$\rm G_{yr}$}
\label{tab:Sun}
\resizebox{\columnwidth}{!}{%
\begin{tabular}{cccccc}
\hline\hline
             && $M$ (M$_\odot$) & $R$ (R$_\odot$) & $\tau$ (Gyr)  & $\rho$ (g cm$^{-3}$) \\ \hline
\multirow{2}{*}{Grid A}& 3:3 & $0.999\pm0.007$ & $0.998\pm0.002$ & $4.68\pm0.22$ & $1.416\pm0.001$      \\
& 3:N & $1.007\pm0.002$ & $1.000\pm0.001$ & $4.66\pm0.06$ & $1.418\pm0.001$      \\ \hline
\multirow{2}{*}{Grid B}& 3:3    & $0.998\pm0.007$ & $0.998\pm0.002$ & $4.69\pm0.22$  & $1.416\pm0.001$      \\
& 3:N.    & $1.007\pm0.002$ & $1.000\pm0.001$ & $4.66\pm0.06$ & $1.418\pm0.001$      \\ \hline
\end{tabular}%
}
\end{table}

\begin{figure}[ht!]
    \centering
    \includegraphics[width=0.97\columnwidth]{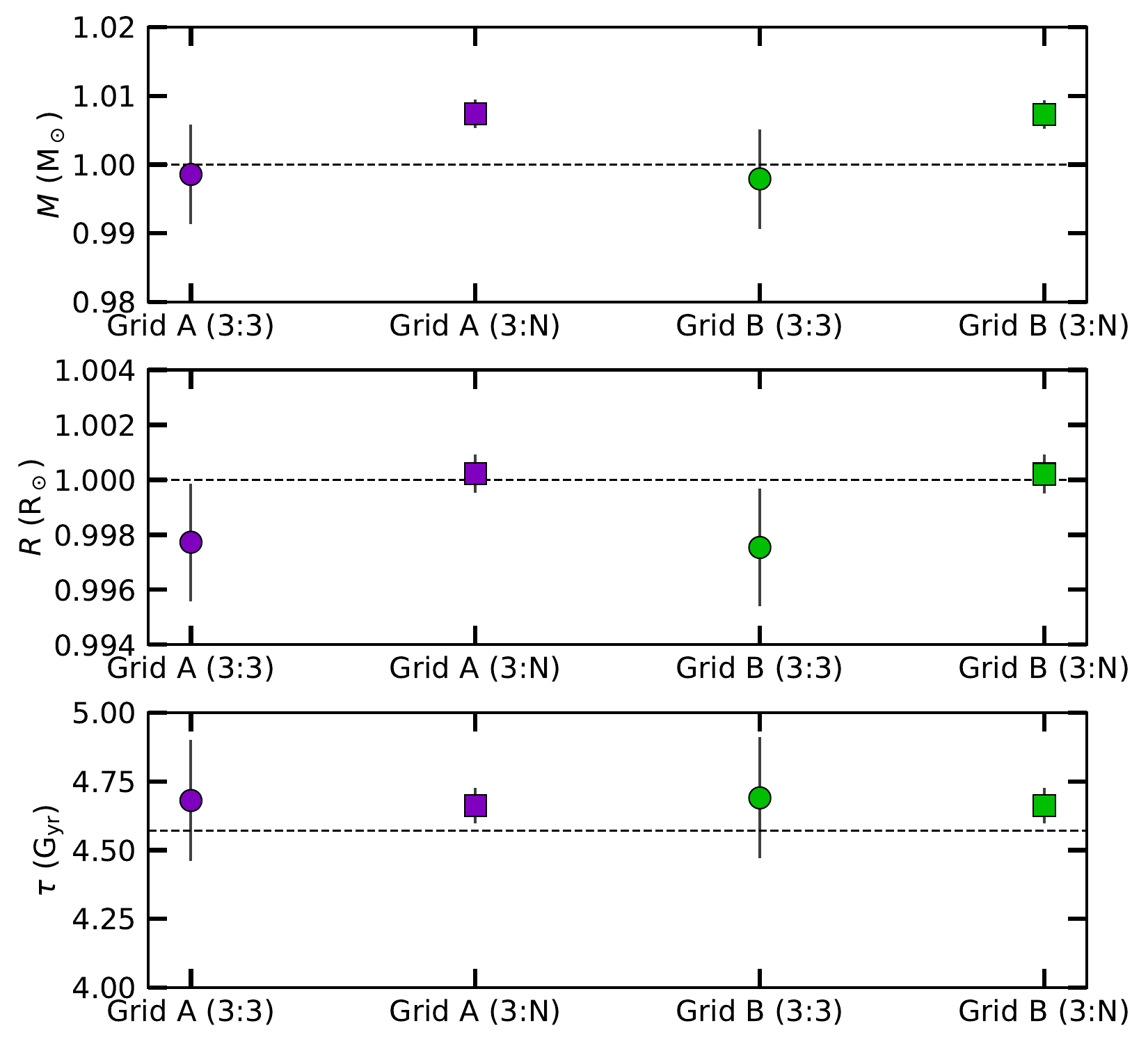}
    \caption{Optimisation results for the Sun, for mass (top panel), radius  (middle panel) and age (bottom panel). The dashed line indicates the real value of the Sun.}
    \label{fig:Sun_res}
\end{figure}

As a first test, we determined how both grids perform in the inference of the properties of the degraded Sun. This tests the accuracy of the grids in the optimisation process. The results for the Sun are shown in Table \ref{tab:Sun} and in Fig. \ref{fig:Sun_res} for both grids with the consideration of the relative or absolute weights in the frequencies.

We see the results are similar for both grids when we use the same weights in the optimisation. This is expected because, for the low-mass models of the grids, the physics is the same. In the models of Grid B, the effect of turbulent mixing is negligible since the convective envelope already fully homogenises the region where it should have an impact. For grid A, most of the models include atomic diffusion.
Comparing the same grid but different weights in the frequencies we see with the 3:3 weights that the true properties of the Sun are within the $1\sigma$ uncertainties, except for the radius which is within $2\sigma$. In contrast, for the 3:N weights only the radius has the true value within $1\sigma$. For the age, it is within $2\sigma$ and for the mass, it is within $4\sigma$.
From a global perspective, the results are compatible with the current Sun and are in agreement with those obtained by \cite{Silva2017}.

\begin{figure}
    \centering
    \includegraphics[width=0.97\columnwidth]{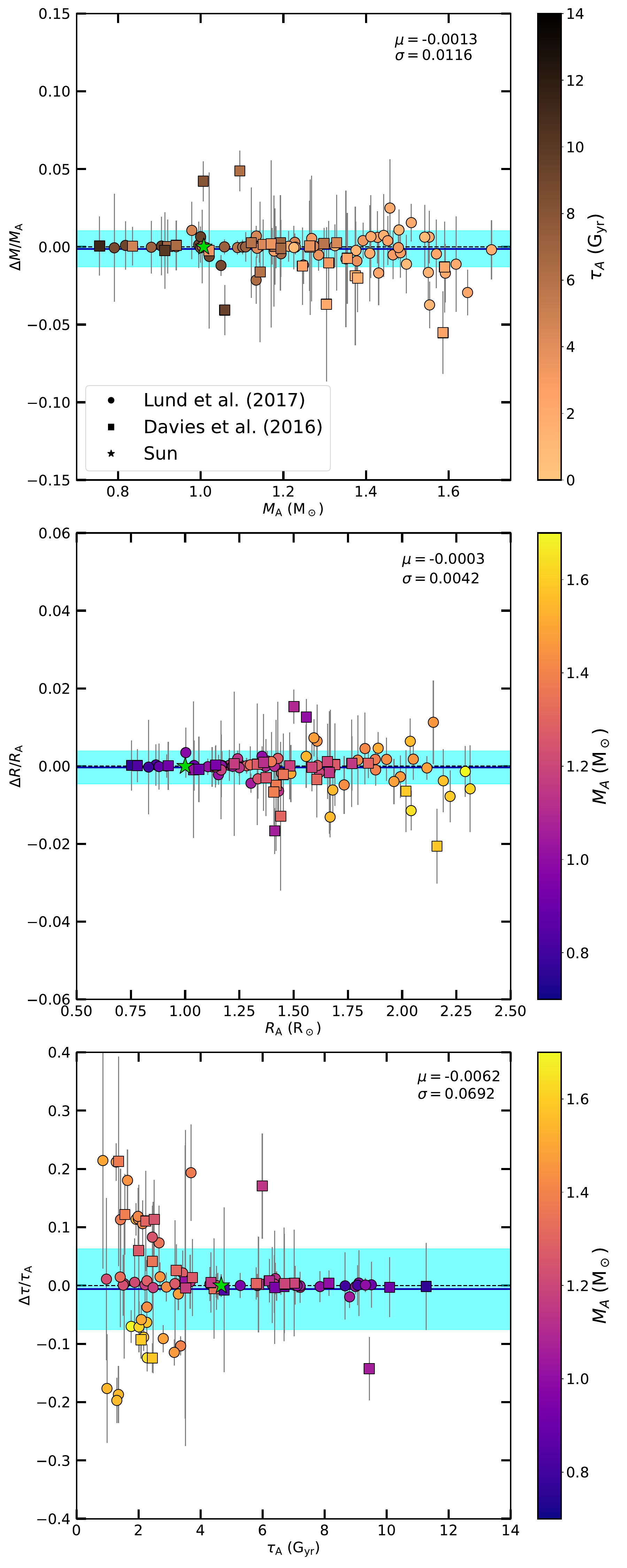}
    \caption{Relative difference for mass (top panel), radius (middle panel) and age (bottom panel) between grid A and B, for 3:N weights. The blue solid line indicates the bias, and the blue region is the one $\sigma$ of the standard deviation. Each point is colour coded with the corresponding reference age (top panel) and mass (middle and bottom panels).}
    \label{fig:stellar_prop_abs_reldiff}
\end{figure}

\begin{figure}
    \centering
    \includegraphics[width=0.97\columnwidth]{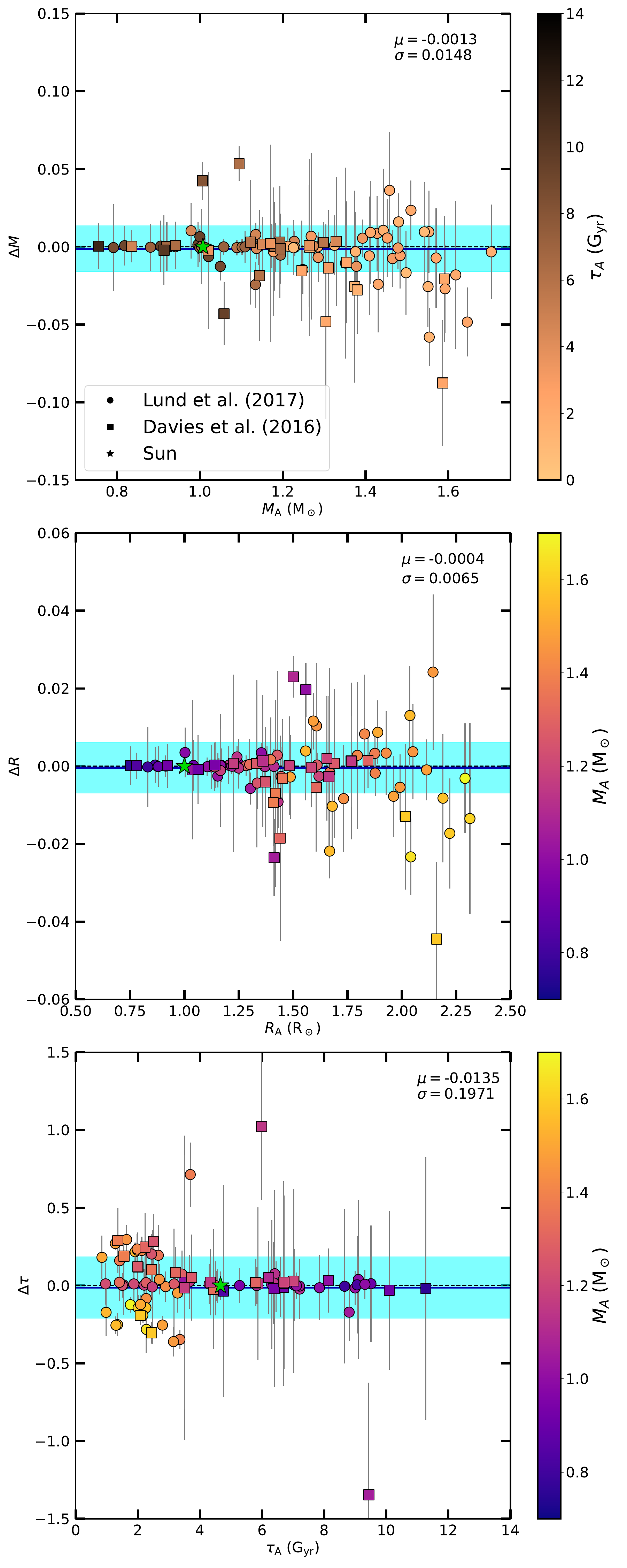}
    \caption{Absolute difference for mass (top panel), radius (middle panel) and age (bottom panel) between grid A and B, for 3:N weights. The blue solid line indicates the bias, and the blue region is the one $\sigma$ of the standard deviation. Each point is colour coded with the corresponding reference age (top panel) and mass (middle and bottom panels).}
    \label{fig:stellar_prop_abs_absdiff}

\end{figure}

\subsection{Grid comparison}

In order to understand how the change of physics (adding turbulent mixing and atomic diffusion for the more massive stars) affects the stellar characterisation, we compare the relative and absolute differences between the fundamental properties inferred with the two grids. In both grids, the inference is made using the 3:N weights of the frequencies in the optimisation because that is the statistically correct option. We estimate the absolute difference with
\begin{equation}
    \Delta X = X_B - X_A,
\end{equation}
and the relative difference with
\begin{equation}
    \frac{\Delta X}{X_A}.
\end{equation}
Where X is the parameter value that is inferred, $X_B$ is the value obtained from grid B and $X_A$ is the reference value obtained from grid A. Figures \ref{fig:stellar_prop_abs_reldiff} and \ref{fig:stellar_prop_abs_absdiff} show the results for the relative and absolute differences for mass, radius, and age, respectively.

Globally, we see that the change in physics does not induce a significant bias in the results. Nevertheless, the mean dispersion for age can reach values of about 7\%,  and the maximum dispersion can reach values greater than 20\% indicating that the changes in physics have a strong impact on the age determination for individual stars.

We also see that the dispersion in all parameters increases as the stellar mass increases, indicating the change we made to the grid has indeed impacted the more massive stars. 
For the lower mass stars both grids include atomic diffusion and the turbulent mixing prescription has not a significant impact (i.e. the turbulent mixing effects are within the convective envelope of these stars).

For the more massive stars, the majority of the models of grid A do not include atomic diffusion.
We expect the models with atomic diffusion (grid B) to be affected in two ways. First, the models have a different surface composition at a given age which has an impact on the inferred stellar properties through the [Fe/H] constraint. Second, atomic diffusion changes the time that the star spends on the main sequence by helping to deplete some of the hydrogen in the core. We see that this can lead to a relative difference for individual stars larger than 5\% in mass, 2\% in radius and 20\% in age.

We look also at the absolute differences (Fig.~\ref{fig:stellar_prop_abs_absdiff}) to verify if the dispersion we see in relative differences at smaller ages is not an artefact caused by the computation of a ratio (particularly age, the small reference value in the denominator may misslead the interpretation). We found the same behaviour in the dispersion of the absolute difference, where it increases with stellar mass. The age dispersion can reach differences of more than 0.4~Gyr for the more massive stars, with much larger differences than those found for the lower masses. This leads us to the conclusion that the dispersion in the relative difference is not fully explained by the smaller ages.

In both the relative and absolute cases, we found 3 outliers for mass and age in low-mass stars. The large difference in these stars is caused by the way Grid A was computed. They encounter the border where atomic diffusion is turned off. This creates a discontinuity in the parameter space where models have and do not have atomic diffusion, which disperses more the results for these stars. This reveals that the current way of grid base modelling where atomic diffusion is cut at a certain point of the grid, to avoid chemical surface over-variations, will be a source of large uncertainties. The most physically consistent approach is to consider atomic diffusion in all the stellar models with the consideration of the other chemical transport mechanisms in competition.

%\begin{figure}[h]
%    \centering
 %   \includegraphics[width=1\columnwidth]{images/Dturb_vs_no_diff/rel/Dturb_vs_no_Dturb_rel__LEG_denci.pdf}
%    \caption{Relative difference of the obtained mean density between grid A and B. The blue solid line indicates the obtained bias, and the blue region is the one $\sigma$ deviation of our sample. Each point is colour coded with the correspondent reference mass that we obtained.}
%    \label{fig:stellar_prop}
%\end{figure}

\section{Discussion}
\label{sec:discussion}

\begin{figure}[ht!]
    \centering
    \includegraphics[width=0.98\columnwidth]{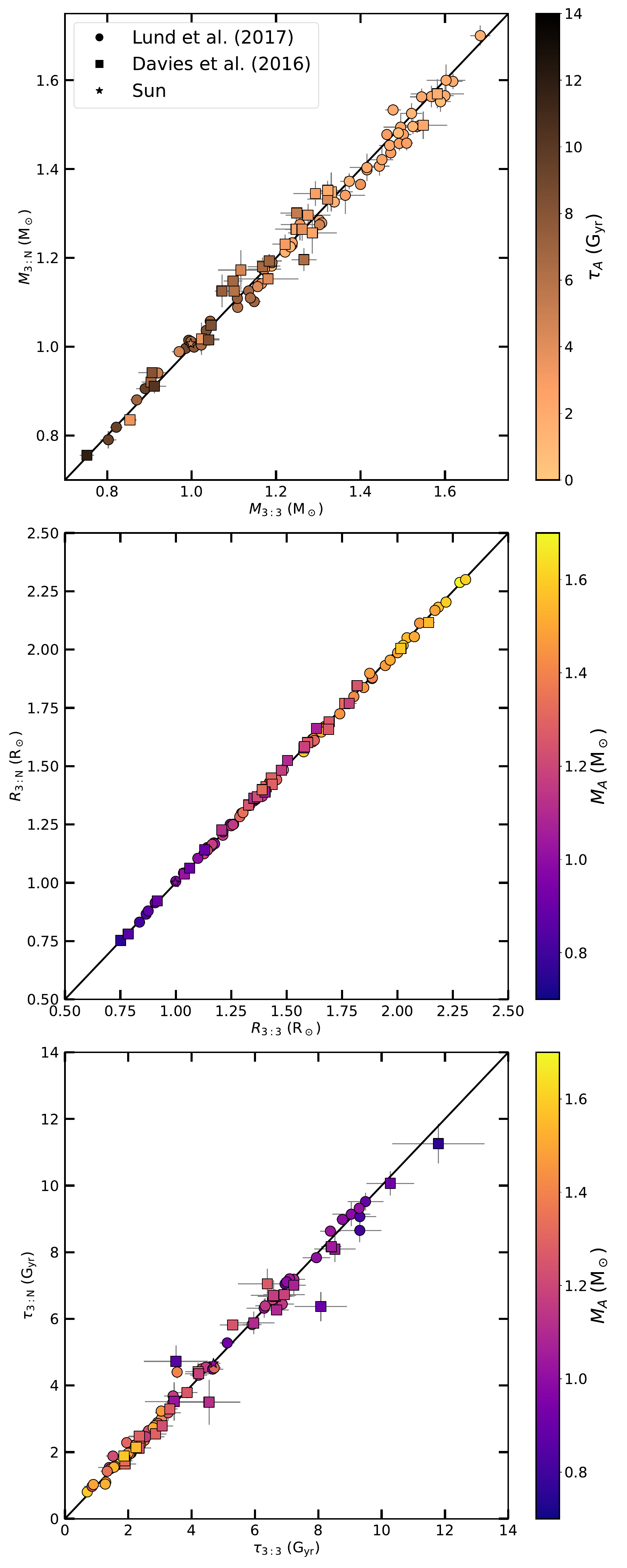}
    \caption{Comparison of the properties inferred when using 3:N or 3:3 weights on the full sample, for mass (top panel), radius  (middle panel) and age (bottom panel). Each point is colour coded with the corresponding reference age (top panel) and mass (middle and bottom panels).}
    \label{fig:abs_vs_rel}
\end{figure}

\begin{figure}[ht!]
    \centering
    \includegraphics[width=0.98\columnwidth]{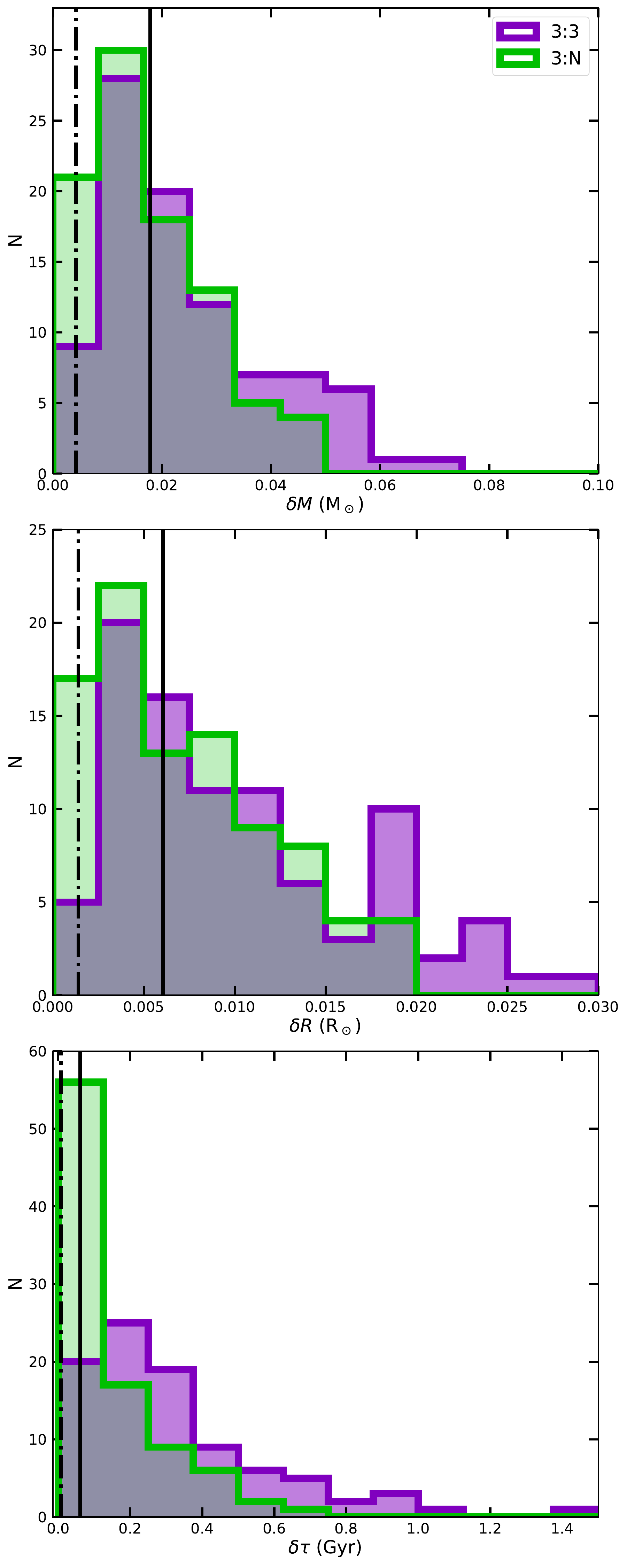}
    \caption{Histogram of uncertainties on the inferred parameters for the full sample and two different weights: mass (top panel), radius  (middle panel) and age (bottom panel). The vertical dash-dotted line marks the value of the bias and the vertical solid line marks the value of the standard deviation.}
    \label{fig:hist_abs_vs_rel}
\end{figure}

\begin{figure}[ht!]
    \centering
    \includegraphics[width=0.98\columnwidth]{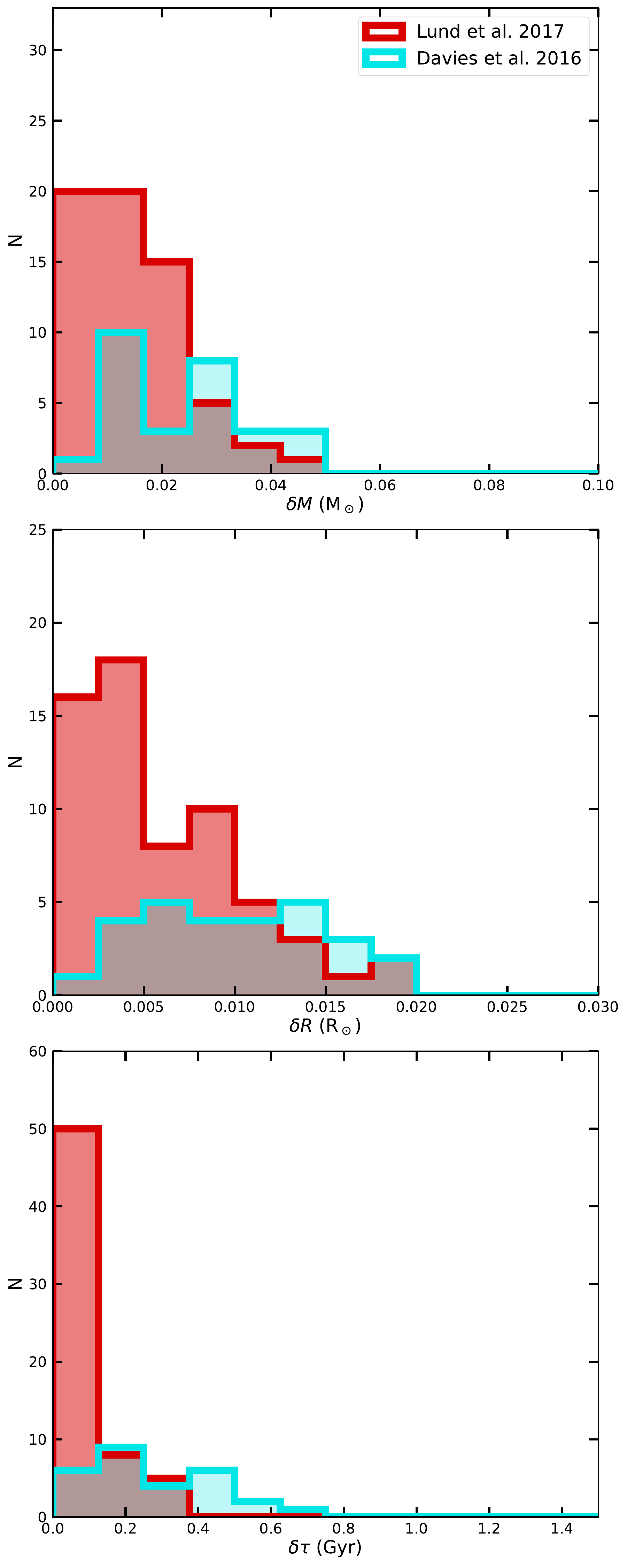}
    \caption{Histogram of uncertainties on the fundamental properties of stars from both samples using the 3:N weights: mass (top panel), radius (middle panel) and age (bottom panel).}
    \label{fig:hist_LEG_vs_DAV}
\end{figure}

\subsection{Impact of  the weight of the frequencies on the inferences}
\label{sec:abs_vs_rel}

In this section, only grid B is used because both grids give similar results when comparing the inference using 3:N and 3:3 frequency weight. Figure \ref{fig:abs_vs_rel} shows the comparison of values inferred using Grid B for the two weight options. The results show that the impact of changing the weights is more significant for the mass and age (top and bottom panels)
%a higher difference in the inference of mass and age depending if we are using the 3:3 or the 3:N results (top and bottom panels).

The parameters uncertainties for both cases are presented in the histograms of Fig. \ref{fig:hist_abs_vs_rel} for all stars. We found that the two methods give different uncertainty distributions, in accordance with \cite{Cunha2021}, because the use of 3:3 weights is similar to synthetically inflating the uncertainties of the frequencies, as expected. We look at the statistics of the relative differences for the inferred mass, radius and age between the two cases. The vertical lines in the histograms show the values of the bias (dash-dotted line) and the standard deviation (solid line). There is no significant bias for any of the parameters, all biases are lower than 1\%. There is a large scatter in the results of 1.8\%, 0.6\% and 6.1\% for mass, radius and age respectively. 

\subsection{Uncertainties comparison between L17 and  D16}
\label{sec:unc_LEG_DAV}
In this section we compare how the quality of the seismic data affects the results, in particular the relationship between the precision of the frequencies and the inferred fundamental properties of the stars. Both L17 and D16 used the same method of seismic identification and quality assurance. The difference is that L17 has the stars with the highest signal-to-noise ratio observed by Kepler, so better seismic quality is expected from this sample. 

In figure \ref{fig:hist_LEG_vs_DAV} we show the histogram of the uncertainties for the two sets of stars regarding mass, radius, and age. D16 sample has a more scattered histogram than the L17 sample and tends to have higher uncertainties, as expected.

There are three stars in common between the samples (KIC3632418, KIC9414417 and KIC10963065), which can be used to see how the different frequency estimates affect the inference of the results (we used the same $\Teff$ and $\FeH$ in the inference, only changing $\nu_\mathrm{max}$ to be consistent with the individual frequency estimate in each work). The results obtained from the seismic data provided by D16 and L17 are presented in table \ref{tab:same_star}. For the three stars in common, D16 data tend to give smaller values of mass, radius and age compared to L17 data, and also show a higher uncertainty for these 3 properties. These differences show that even a small change in the estimated individual frequencies can have a significant impact on the result. Something that is somewhat surprising is that lower masses have lower ages in these three stars from the D16 sample. A possible explanation is the difference in the initial chemical compositions ($\MH_i$ and $Y_i$). For the case of D16, these three models of these stars have slightly lower $\MH_i$ and higher $Y_i$, which may cause the age to be smaller for lower estimated masses.

Uncertainties are expected to decrease with improved quality. This was tested on synthetic stars in the work of \cite{Cunha2021}, who found that the degraded data gave less accurate results, especially for age determination. This is in agreement with what we found for the \textit{Kepler} data we analysed.

\begin{table*}
\centering
\caption{Inferred properties of the stars that are common to the D16 and L17 samples using the individual frequencies obtained in the respective works.}
\label{tab:same_star}
\resizebox{\textwidth}{!}{%
\begin{tabular}{cccccccccccc}

\hline\hline
\multicolumn{2}{c}{\multirow{2}{*}{KIC}} & \multicolumn{2}{c}{Mass ($\Msun$)} & \multicolumn{2}{c}{Radius (R$_\odot$)} & \multicolumn{2}{c}{Age ($\rm G_{yr}$)} & \multicolumn{2}{c}{$\boldsymbol{\MH_i}$} & \multicolumn{2}{c}{$\boldsymbol{Y_i}$} \\ %\cline{3-8} 
\multicolumn{2}{c}{} & D16 & L17 & D16 & L17 & D16 & L17 & D16 & L17 & D16 & L17 \\ \hline
\multirow{2}{*}{3632418} & (3:3) & $1.420\pm0.042$ & $1.471\pm0.013$ & $1.917\pm0.020$ & $1.945\pm0.006$ & $2.795\pm0.127$ & $2.927\pm0.121$ &$\boldsymbol{0.159\pm0.049}$ & $\boldsymbol{0.171\pm0.054}$& $\boldsymbol{0.278\pm0.024}$& $\boldsymbol{0.271\pm0.022}$\\

 & (3:N) & $1.405\pm0.024$ & $1.437\pm0.017$ & $1.911\pm0.011$ & $1.931\pm0.008$ & $2.719\pm0.051$ & $2.888\pm0.051$ &$\boldsymbol{0.221\pm0.023}$ & $\boldsymbol{0.228\pm0.026}$& $\boldsymbol{0.290\pm0.013}$& $\boldsymbol{0.289\pm0.014}$\\ \hline
 
\multirow{2}{*}{9414417} & (3:3) & $1.434\pm0.052$ & $1.502\pm0.019$ & $1.935\pm0.025$ & $1.967\pm0.009$ & $2.483\pm0.125$ & $2.571\pm0.103$ & $\boldsymbol{0.151\pm0.052}$ & $\boldsymbol{0.152\pm0.068}$& $\boldsymbol{0.275\pm0.070}$ & $\boldsymbol{0.276\pm0.024}$\\

 & (3:N) & $1.455\pm0.035$ & $1.478\pm0.017$ & $1.944\pm0.016$ & $1.955\pm0.008$ & $2.380\pm0.057$ & $2.539\pm0.048$ & $\boldsymbol{0.214\pm0.038}$ & $\boldsymbol{0.217\pm0.040}$& $\boldsymbol{0.281\pm0.016}$ & $\boldsymbol{0.282\pm0.040}$\\
 
\hline
\multirow{2}{*}{10963065} & (3:3) & $1.111\pm0.027$ & $1.156\pm0.009$ & $1.239\pm0.011$ & $1.258\pm0.003$ & $4.211\pm0.319$ & $4.223\pm0.267$ & $\boldsymbol{0.016\pm0.060}$ &$\boldsymbol{0.043\pm0.062}$& $\boldsymbol{0.262\pm0.017}$ & $\boldsymbol{0.264\pm0.017}$\\

 & (3:N) & $1.108\pm0.013$ & $1.135\pm0.005$ & $1.237\pm0.005$ & $1.249\pm0.002$ & $4.121\pm0.095$ & $4.307\pm0.089$ & $\boldsymbol{-0.044\pm0.043}$ &$\boldsymbol{-0.054\pm0.044}$& $\boldsymbol{0.258\pm0.011}$&$\boldsymbol{0.259\pm0.011}$\\
 \hline

\end{tabular}%
}
\end{table*}

\begin{figure}[ht!]
    \centering
    \includegraphics[width=0.95\columnwidth]{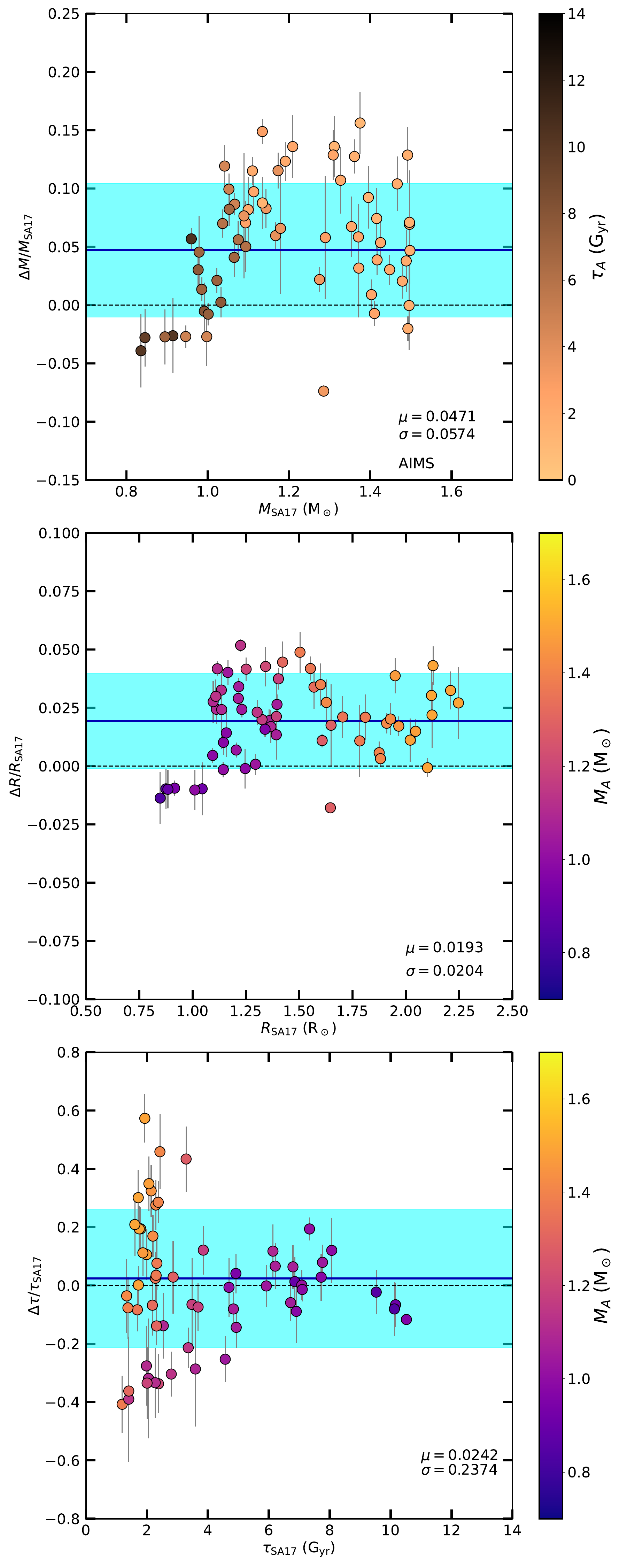}
    \caption{Relative difference for mass (top panel), radius (middle panel) and age (bottom panel) between grid B and \cite{Silva2017} results for the AIMS pipeline. The blue solid line indicates the bias, and the blue region is the one $\sigma$ deviation. Each point is colour coded with the corresponding reference age (top panel) and mass (middle and bottom panels).}
    \label{fig:r_vs_SA17_AIMS}
\end{figure}

\begin{figure}[ht!]
    \centering
    \includegraphics[width=0.95\columnwidth]{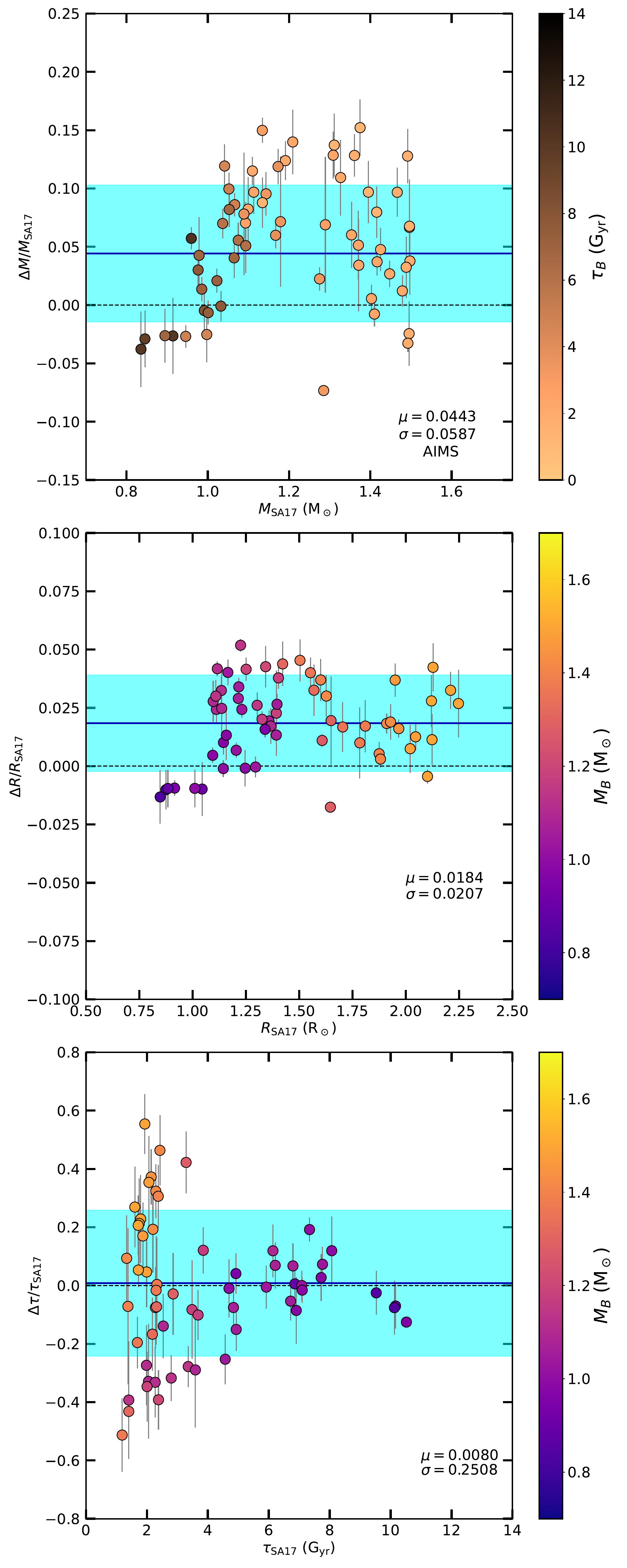}
    \caption{Same as Fig. \ref{fig:r_vs_SA17_AIMS} but using our results from Grid A.}
    \label{fig:r_vs_SA17_AIMS_grid_a}
\end{figure}

\subsection{Comparison with results of \cite{Silva2015} and \cite{Silva2017}}

In this section, we compare the results obtained using Grid B with previous work. We compare our study of the L17 sample with the work of \cite{Silva2017} and our study of the D16 sample with the work of \cite{Silva2015}. In both works, they use different pipelines and the grids were} built with different codes and input physics. They also use different optimisation codes in each pipeline and use a relative weight for each frequency. Here we compare their results with our results derived using the same relative weight (for consistency).

\subsubsection{Comparison with \cite{Silva2017}}

We compare our results for the L17 sample with the results presented in \cite{Silva2017}. We show here the results for only one of the seven pipelines used in their paper, i.e. ``AIMS''. Similar conclusions can be drawn for all the others. This pipeline uses the same evolutionary code (MESA) to compute the stellar models as in our work, with different physical inputs, more precisely, use \cite{GN93} solar mixture, Eddington gray atmosphere, no atomic diffusion, and a fix helium enrichment ratio ($\frac{\Delta Y}{\Delta Z}=2.0$).

Figure \ref{fig:r_vs_SA17_AIMS} shows the comparison of Grid B results with those from the AIMS pipeline of \cite{Silva2017}.  Our results show a bias towards higher values for mass, radius and age, with a scatter of 6\%, 2\% and 24\% respectively. However, the relative difference for an individual star can be up to 15\% for mass, 5\% for radius and 60\% for age.

We also compared the results from \cite{Silva2017} with those of grid A in Fig. \ref{fig:r_vs_SA17_AIMS_grid_a}; however, for most cases, we find similar significant bias and scatter as in the comparison with grid B. This suggests that the systematic effects arising from turbulent mixing are overshadowed by the different physics used between our work and \cite{Silva2017}.

\subsubsection{Comparison with \cite{Silva2015}}

\begin{figure}
    \centering
    \includegraphics[width=0.95\columnwidth]{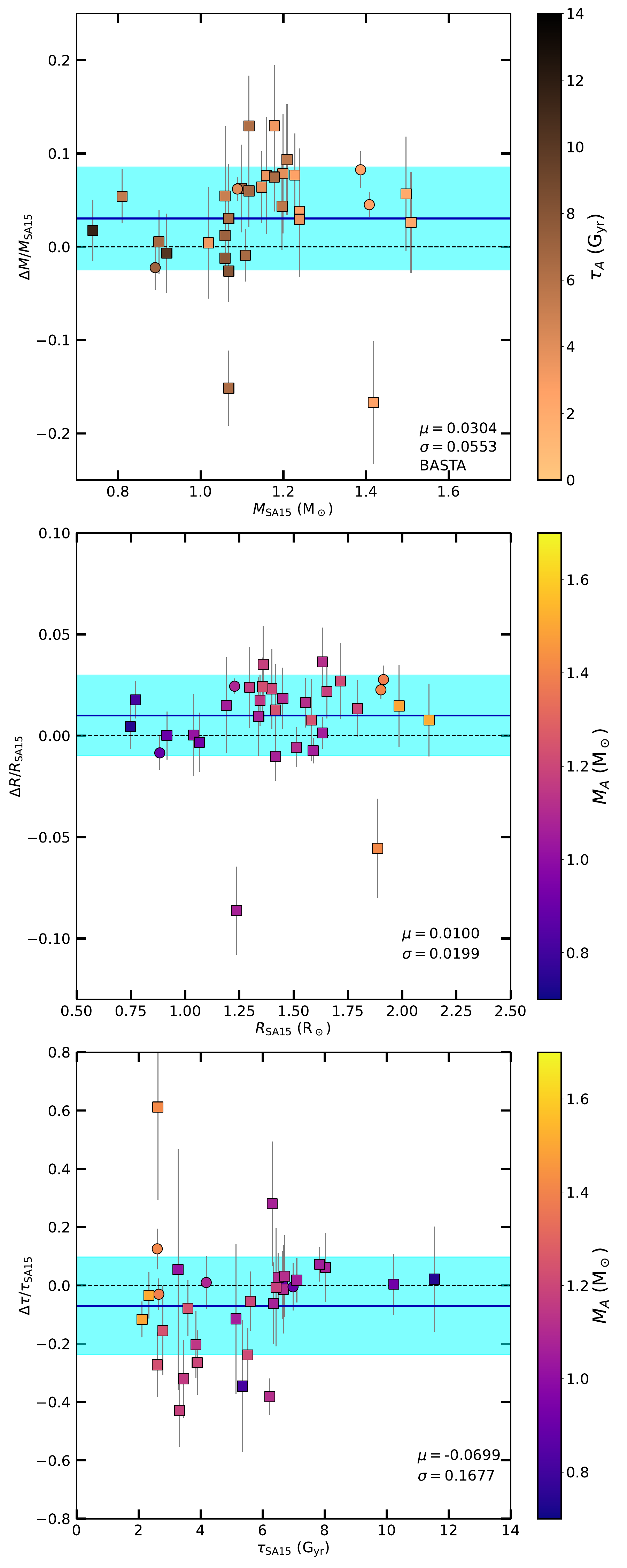}
    \caption{Relative difference for mass (top panel), radius (middle panel) and age (bottom panel) between grid B and \cite{Silva2015} results for the BASTA pipeline. The blue solid line indicates the bias, and the blue region is the one $\sigma$ deviation. Each point is colour-coded with the corresponding reference age (top panel) and mass (middle and bottom panels).}
    \label{fig:r_vs_SA15_BASTA}
\end{figure}

We now compare our results for the stars of D16 with those presented by \cite{Silva2015}. Like in the previous section, the results obtained from all the pipelines in their work were similar and led to the same conclusions. Therefore we only present here the comparison with the BASTA pipeline (Fig \ref{fig:r_vs_SA15_BASTA})

 We observe a bias towards higher masses and radii, and a bias towards younger ages, with a dispersion of 6\%, 2\% and 17\% for mass, radius and age respectively. For some individual stars, the relative difference can be up to 20\% for mass, 10\% for radius and 60\% for age.

\begin{figure}
    \centering
    \includegraphics[width=0.95\columnwidth]{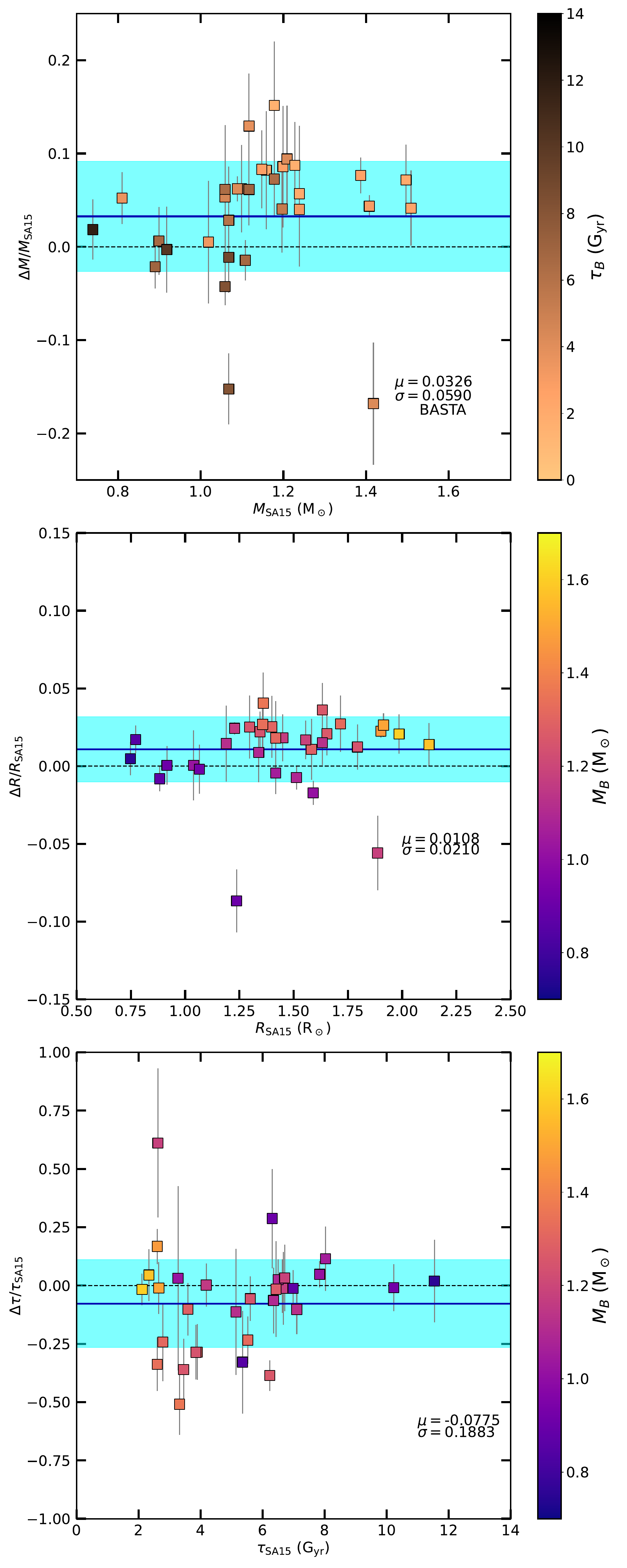}
    \caption{Same as Fig. \ref{fig:r_vs_SA15_BASTA} but using our results from Grid A.}
    \label{fig:r_vs_SA15_BASTA_grid_a}
\end{figure}

As before, we also compared the \cite{Silva2015} with the results from grid A in Fig. \ref{fig:r_vs_SA15_BASTA_grid_a}. However, we identify a similar bias and scatter as in the comparison with grid B. This again points out that the systematics we observe are a consequence of the different physics used between the two works and not specific related to the incorporation of turbulent mixing in our grid B. 

\section{Conclusion}
\label{sec:conclusion}

This work is a continuation of the study by \cite{moedas2022} to understand how turbulent mixing and atomic diffusion affects the stellar characterisation of F-type stars. The main objective was to study the impact on the stellar properties of including calibrated turbulent mixing in stellar models. Including this mechanism allows also the inclusion of atomic diffusion in F-type stars, avoiding surface chemical over-variations. To do this, we computed two grids from which we inferred the stellar properties of a sample of FGK-type stars. The first grid (A) neglects atomic diffusion for the F-type stars, the second grid (B) uses our calibrated turbulent mixing, allowing the use of atomic diffusion in stellar models.

In addition to studying the impact of the turbulent mixing, we tested how applying different weights to the seismic data impacts the results. Furthermore, by selecting samples from two different sources \cite{Davies2016} and \cite{Lund2017} we investigated how the data quality affects the uncertainties of the properties of observed stars. Finally, we compared our results with previous ones that analysed the same samples of stars.
%We carried out other tests, firstly to see how the way in which the different weights in the seismic data are taken into account affects the results. Secondly, we selected the sample from two different sources, D16 and L17, so that we can investigate how the data quality affects the uncertainties of the observed star (L17 represent the stars with the highest signal-to-noise observed by \textit{Kepler}), were also conducted in \cite{Cunha2021} with synthetic stars. Finally, we compare our results with previous ones that analysed the same sample of stars.

Concerning the combined inclusion of turbulent mixing and atomic diffusion, globally speaking we found no significant impact, i.e. a relative bias of less than 1\% for masses, radii and ages. However, we found that there is an increase in the dispersion of the relative differences with mass, caused by neglecting atomic diffusion in F-type stars. This can lead to individual relative differences of up to 5\% for mass, 2\% for radius and 20\% for age. This shows that  including atomic diffusion is necessary if we want to avoid this source of uncertainties. We also found three stars that can be considered outliers. In fact, their best-fit models are in the limit where atomic diffusion is turned off for grid A for F-type stars. This type of discontinuity in the parameter space introduced in grids by considering models with and without diffusion (such as in our grid A) can therefore introduce significant errors in the inferred stellar properties. Our conclusion is that in this region of the parameter space, we need the most homogeneous physics across the grid, avoiding discontinuity problems. This result shows that, in order to reduce the uncertainties we have to consider atomic diffusion in all stellar modes, in combination with other chemical transport mechanisms to avoid unrealistic surface abundance variations. Turbulent mixing is a parameterisation of the different processes and is a step that allows us to improve stellar models and better characterise stars. Although, this improves the prediction of the evolution of iron in stellar models we still need to take into consideration that it may not be appropriated for other elements (i.e. oxygen and calcium).

The results from other tests we performed on the inference method and data quality are consistent with what was found by \cite{Cunha2021} using synthetic stars. The use of a weight of (3:N) instead of (3:3) for the individual frequencies constrains leads to results that are more sensitive to the input physics considered in the stellar models, and to smaller uncertainties, as expected. For the quality test, the better quality in the individual frequencies of L17  provides smaller uncertainties in the inferred properties compared to D16. The results for the three in-common stars (KIC3632418, KIC9414417 and KIC10963065) also lead to a similar conclusion. 
Nonetheless, the small changes in the determination of the seismic data in these three stars lead to small differences in the inferred values of mass, radius, age, and initial chemical composition.

We compared the results of our sample with previous works. For D16 we compared with the results of \cite{Silva2015} and for L17 with \cite{Silva2017}. We found that there are large differences in both cases, due to the different physics adopted in this work compared to the two previously mentioned. This reinforces the need for careful consideration of the input physics.

This work demonstrates that the calibrated turbulent mixing of \cite{moedas2022} allows a better characterisation of observed F-type stars, even if the proposed scheme is not able to reproduce the chemical evolution of all individual elements. In order to overcome this issue, a further step can be the implementation in MESA of the Single Value Parameter (SVP) method \citep{alecian20} which provides a good balance in efficiency between the calculation of the radiative accelerations and the computation time. Nevertheless, this calibrated turbulent mixing is a step towards a better characterisation of stars with more accurate physics (including atomic diffusion). Finally, with this work, we provide an updated characterisation for the fundamental parameters (masses, radii and ages) for the D16 and L17 stars analysed. 

\section*{Acknowledgements}

This work was supported by FCT/MCTES through the research grants UIDB/04434/2020, UIDP/04434/2020 2022.06962.PTDC., 2022.03993.PTDC, and DOI 10.54499/2022.03993.PTDC.  NM acknowledges support from the Fundação para a Ciência e a Tecnologia (FCT) through the Fellowship UI/BD/152075/2021 and POCH/FSE (EC). DB is supported by national funds through FCT in the form of a work contract. MC acknowledges the support by national funds (FCT/MCTES, Portugal), through the contract CEECIND/02619/2017.
We also thank to Daniel Reese for providing us with the results from AIMS pipeline from \cite{Silva2017}. We thank the anonymous referee   for the valuable comments which helped to improve the paper.\\ 

%-------------------------------------------------------------------------------------------------------------
%-------------------------------------------------------------------------------------------------------------
%-------------------------------------------------------------------------------------------------------------
%-------------------------------------------------------------------------------------------------------------
%-------------------------------------------------------------------------------------------------------------
%-------------------------------------------------------------------------------------------------------------
%-------------------------------------------------------------------------------------------------------------

\bibliographystyle{aa} % style aa.bst
\bibliography{Version_f.bib} % your references Yourfile.bib

%\newpage

\begin{appendix}

\newpage
\FloatBarrier

\section{Grid comparison 3:3 frequency weight}

\begin{figure}[ht!]
    \centering
    \includegraphics[width=0.95\columnwidth]{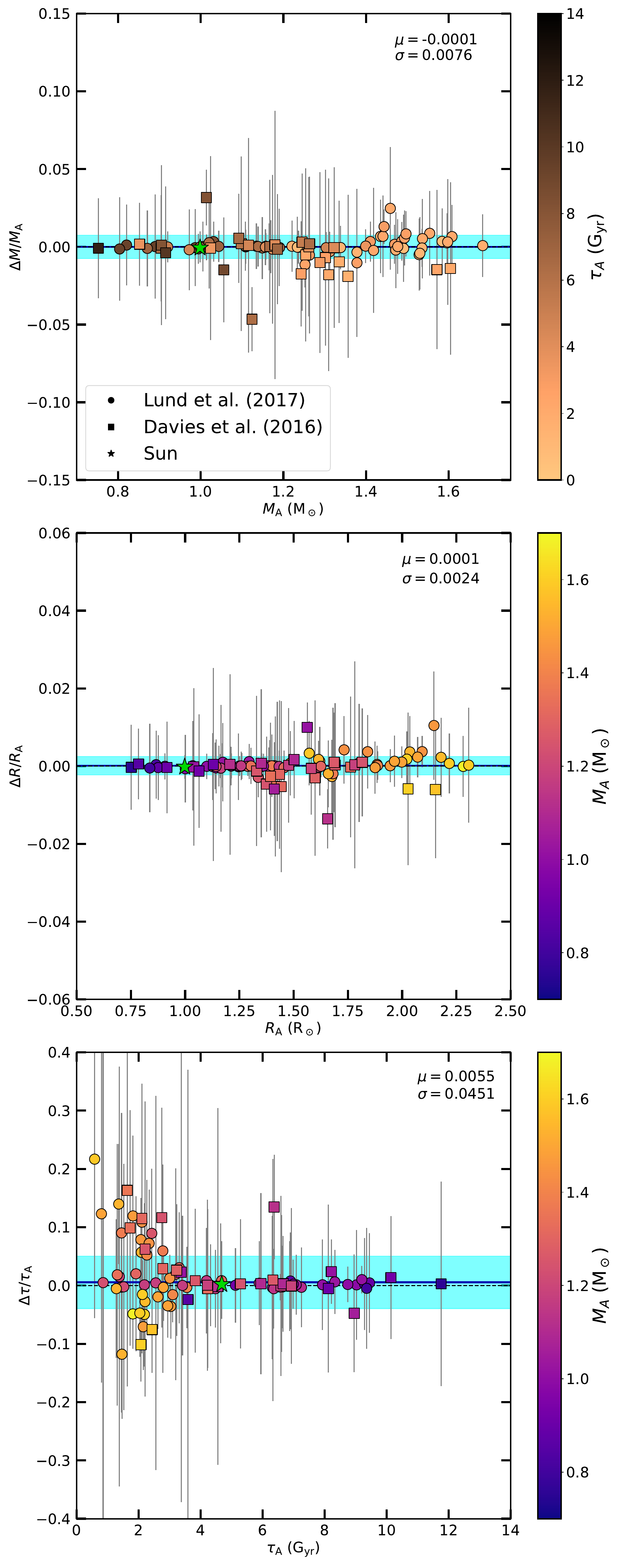}
    \caption{ Same as Fig. \ref{fig:stellar_prop_abs_reldiff} but using 3:3 weights for frequencies.}
    \label{fig:stellar_prop_rel_reldiff}
\end{figure}

\begin{figure}[ht!]
    \centering
    \includegraphics[width=0.95\columnwidth]{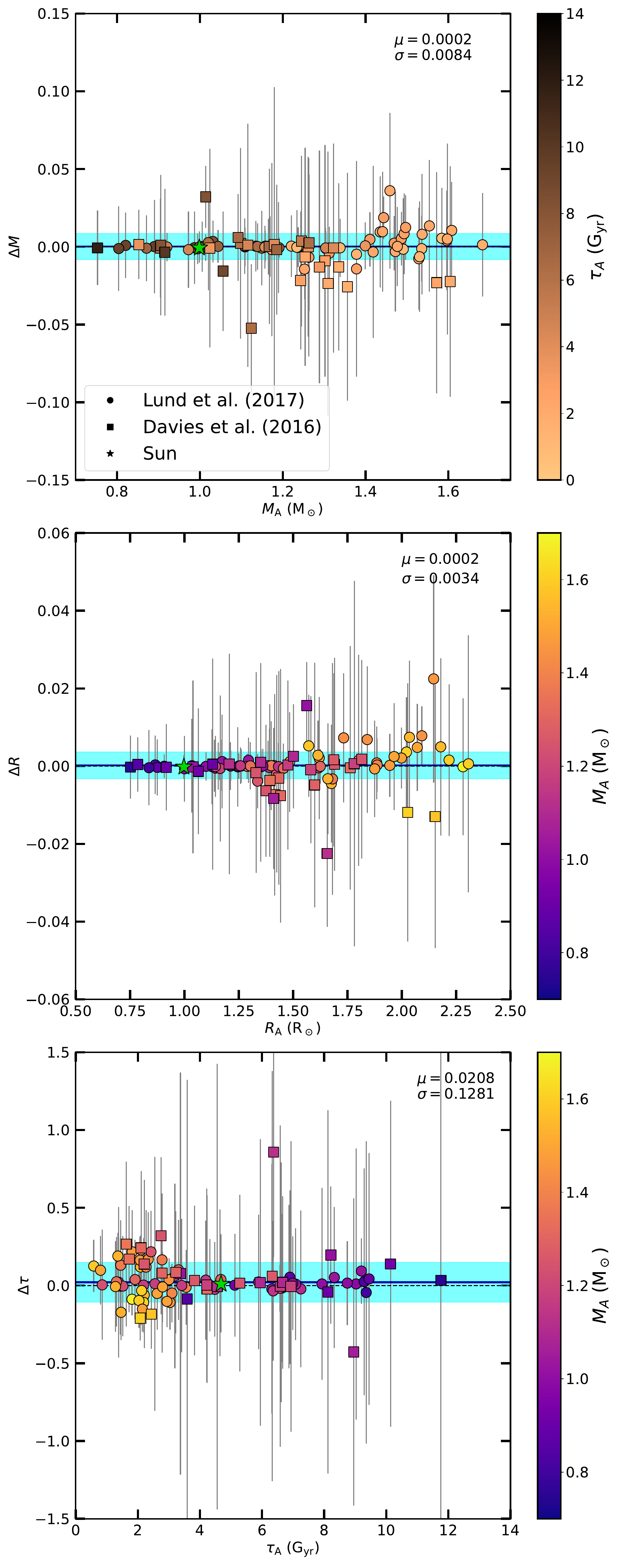}
    \caption{Same as Fig. \ref{fig:stellar_prop_abs_absdiff} but using 3:3 weights for frequencies}
    \label{fig:stellar_prop_rel_absdiff}
\end{figure}

Figures \ref{fig:stellar_prop_rel_reldiff} and \ref{fig:stellar_prop_rel_absdiff} are the same as Fig. \ref{fig:stellar_prop_abs_reldiff} and \ref{fig:stellar_prop_abs_absdiff} but for the 3:3 weights. In this case, we can see the same behaviour as in the case of the 3:N weights. However, we can see that for the case of relative weights, there is a smaller dispersion (which we can see in the standard deviation). This is due to the fact that the use of absolute weights is more sensitive to the input physics. Nevertheless, the conclusion we obtained using the 3:N or 3:3 weights in frequencies is the same, but more pronounced for the 3:N weights.

\newpage
\FloatBarrier
\section{Fundamental properties using 3:N frequencies weight}

\onecolumn
\longtab{
\begin{longtable}{cccccccccccc}
\caption{Inferred fundamental properties of the stars using 3:N frequencies weight.}
\label{tab:abs_res_3N}
\\\hline
KIC       & Mass (M$_\odot$) & Radius (R$_\odot$) & Age (Gyr) &  $\MH_i$ &    $Y_i$ \\
\hline
\endfirsthead
\caption{continued}\\
\hline
KIC       & Mass (M$_\odot$) & Radius (R$_\odot$) & Age (Gyr) &  $\MH_i$ &    $Y_i$ \\
\hline
\endhead
\hline
\endfoot
\hline
\endlastfoot
1435467   & 1.406 $\pm$ 0.021 & 1.724 $\pm$ 0.009 & 2.184 $\pm$ 0.044 & 0.251 $\pm$ 0.045 & 0.301 $\pm$ 0.013 &\\
2837475   & 1.525 $\pm$ 0.023 & 1.670 $\pm$ 0.009 & 1.096 $\pm$ 0.051 & 0.279 $\pm$ 0.047 & 0.289 $\pm$ 0.016 &\\
3427720   & 1.174 $\pm$ 0.010 & 1.140 $\pm$ 0.004 & 2.205 $\pm$ 0.075 & 0.043 $\pm$ 0.039 & 0.243 $\pm$ 0.003 &\\
3456181   & 1.563 $\pm$ 0.025 & 2.181 $\pm$ 0.013 & 2.117 $\pm$ 0.045 & 0.042 $\pm$ 0.050 & 0.258 $\pm$ 0.012 &\\
3632418   & 1.437 $\pm$ 0.017 & 1.931 $\pm$ 0.008 & 2.888 $\pm$ 0.051 & 0.213 $\pm$ 0.035 & 0.283 $\pm$ 0.012 &\\
3656476   & 1.037 $\pm$ 0.005 & 1.298 $\pm$ 0.002 & 8.634 $\pm$ 0.121 & 0.287 $\pm$ 0.007 & 0.306 $\pm$ 0.004 &\\
3735871   & 1.182 $\pm$ 0.011 & 1.124 $\pm$ 0.004 & 1.538 $\pm$ 0.139 & 0.045 $\pm$ 0.041 & 0.245 $\pm$ 0.005 &\\
4914923   & 1.126 $\pm$ 0.006 & 1.381 $\pm$ 0.003 & 6.487 $\pm$ 0.122 & 0.179 $\pm$ 0.025 & 0.280 $\pm$ 0.005 &\\
5184732   & 1.231 $\pm$ 0.005 & 1.351 $\pm$ 0.002 & 4.499 $\pm$ 0.065 & 0.415 $\pm$ 0.005 & 0.285 $\pm$ 0.003 &\\
5773345   & 1.563 $\pm$ 0.018 & 2.051 $\pm$ 0.008 & 2.131 $\pm$ 0.036 & 0.399 $\pm$ 0.016 & 0.298 $\pm$ 0.008 &\\
5950854   & 0.996 $\pm$ 0.011 & 1.250 $\pm$ 0.005 & 9.141 $\pm$ 0.362 & -0.103 $\pm$ 0.049 & 0.260 $\pm$ 0.012 &\\
6106415   & 1.143 $\pm$ 0.005 & 1.244 $\pm$ 0.002 & 4.547 $\pm$ 0.074 & 0.049 $\pm$ 0.027 & 0.257 $\pm$ 0.006 &\\
6116048   & 1.089 $\pm$ 0.004 & 1.251 $\pm$ 0.001 & 5.828 $\pm$ 0.082 & -0.153 $\pm$ 0.020 & 0.244 $\pm$ 0.003 &\\
6225718   & 1.284 $\pm$ 0.005 & 1.283 $\pm$ 0.002 & 2.285 $\pm$ 0.057 & 0.182 $\pm$ 0.019 & 0.243 $\pm$ 0.003 &\\
6508366   & 1.565 $\pm$ 0.021 & 2.204 $\pm$ 0.010 & 1.970 $\pm$ 0.037 & 0.243 $\pm$ 0.053 & 0.297 $\pm$ 0.014 &\\
6603624   & 1.015 $\pm$ 0.005 & 1.151 $\pm$ 0.002 & 8.982 $\pm$ 0.086 & 0.219 $\pm$ 0.020 & 0.283 $\pm$ 0.004 &\\
6679371   & 1.700 $\pm$ 0.023 & 2.288 $\pm$ 0.011 & 1.637 $\pm$ 0.036 & 0.279 $\pm$ 0.033 & 0.277 $\pm$ 0.013 &\\
6933899   & 1.189 $\pm$ 0.004 & 1.616 $\pm$ 0.002 & 6.579 $\pm$ 0.067 & 0.061 $\pm$ 0.009 & 0.250 $\pm$ 0.002 &\\
7103006   & 1.597 $\pm$ 0.017 & 2.018 $\pm$ 0.007 & 1.993 $\pm$ 0.040 & 0.389 $\pm$ 0.028 & 0.285 $\pm$ 0.011 &\\
7206837   & 1.453 $\pm$ 0.027 & 1.618 $\pm$ 0.011 & 1.936 $\pm$ 0.056 & 0.368 $\pm$ 0.033 & 0.282 $\pm$ 0.014 &\\
7296438   & 1.102 $\pm$ 0.004 & 1.371 $\pm$ 0.002 & 6.443 $\pm$ 0.115 & 0.292 $\pm$ 0.019 & 0.308 $\pm$ 0.003 &\\
7510397   & 1.398 $\pm$ 0.011 & 1.876 $\pm$ 0.005 & 3.008 $\pm$ 0.041 & -0.041 $\pm$ 0.025 & 0.252 $\pm$ 0.007 &\\
7680114   & 1.109 $\pm$ 0.007 & 1.412 $\pm$ 0.003 & 7.188 $\pm$ 0.135 & 0.158 $\pm$ 0.035 & 0.279 $\pm$ 0.008 &\\
7771282   & 1.341 $\pm$ 0.042 & 1.667 $\pm$ 0.018 & 2.854 $\pm$ 0.116 & 0.070 $\pm$ 0.062 & 0.275 $\pm$ 0.023 &\\
7871531   & 0.819 $\pm$ 0.009 & 0.865 $\pm$ 0.003 & 9.067 $\pm$ 0.220 & -0.107 $\pm$ 0.031 & 0.297 $\pm$ 0.009 &\\
7940546   & 1.458 $\pm$ 0.016 & 1.986 $\pm$ 0.008 & 2.787 $\pm$ 0.053 & 0.170 $\pm$ 0.021 & 0.276 $\pm$ 0.010 &\\
8006161   & 0.941 $\pm$ 0.006 & 0.914 $\pm$ 0.002 & 5.278 $\pm$ 0.080 & 0.334 $\pm$ 0.012 & 0.305 $\pm$ 0.004 &\\
8150065   & 1.275 $\pm$ 0.036 & 1.430 $\pm$ 0.015 & 3.179 $\pm$ 0.243 & 0.123 $\pm$ 0.062 & 0.265 $\pm$ 0.020 &\\
8179536   & 1.373 $\pm$ 0.018 & 1.399 $\pm$ 0.008 & 1.574 $\pm$ 0.082 & 0.150 $\pm$ 0.052 & 0.248 $\pm$ 0.007 &\\
8228742   & 1.365 $\pm$ 0.009 & 1.879 $\pm$ 0.005 & 4.402 $\pm$ 0.196 & 0.230 $\pm$ 0.036 & 0.264 $\pm$ 0.006 &\\
8379927   & 1.229 $\pm$ 0.004 & 1.159 $\pm$ 0.002 & 1.501 $\pm$ 0.055 & 0.168 $\pm$ 0.023 & 0.242 $\pm$ 0.002 &\\
8394589   & 1.142 $\pm$ 0.008 & 1.204 $\pm$ 0.003 & 3.686 $\pm$ 0.120 & -0.077 $\pm$ 0.039 & 0.246 $\pm$ 0.005 &\\
8424992   & 0.906 $\pm$ 0.012 & 1.042 $\pm$ 0.005 & 9.519 $\pm$ 0.266 & 0.024 $\pm$ 0.052 & 0.296 $\pm$ 0.014 &\\
8938364   & 1.006 $\pm$ 0.005 & 1.357 $\pm$ 0.002 & 9.321 $\pm$ 0.079 & -0.227 $\pm$ 0.019 & 0.245 $\pm$ 0.003 &\\
9025370   & 0.989 $\pm$ 0.013 & 1.006 $\pm$ 0.005 & 4.491 $\pm$ 0.121 & 0.059 $\pm$ 0.051 & 0.288 $\pm$ 0.012 &\\
9098294   & 0.999 $\pm$ 0.007 & 1.154 $\pm$ 0.003 & 7.836 $\pm$ 0.147 & -0.096 $\pm$ 0.038 & 0.259 $\pm$ 0.009 &\\
9139151   & 1.213 $\pm$ 0.008 & 1.171 $\pm$ 0.003 & 1.881 $\pm$ 0.099 & 0.115 $\pm$ 0.034 & 0.245 $\pm$ 0.005 &\\
9139163   & 1.496 $\pm$ 0.014 & 1.605 $\pm$ 0.005 & 1.546 $\pm$ 0.026 & 0.387 $\pm$ 0.024 & 0.274 $\pm$ 0.007 &\\
9206432   & 1.552 $\pm$ 0.022 & 1.562 $\pm$ 0.009 & 0.805 $\pm$ 0.070 & 0.277 $\pm$ 0.054 & 0.253 $\pm$ 0.009 &\\
9353712   & 1.494 $\pm$ 0.032 & 2.168 $\pm$ 0.016 & 2.216 $\pm$ 0.052 & 0.085 $\pm$ 0.046 & 0.287 $\pm$ 0.018 &\\
9410862   & 1.004 $\pm$ 0.017 & 1.167 $\pm$ 0.007 & 6.319 $\pm$ 0.291 & -0.213 $\pm$ 0.053 & 0.266 $\pm$ 0.015 &\\
9414417   & 1.478 $\pm$ 0.017 & 1.955 $\pm$ 0.008 & 2.539 $\pm$ 0.048 & 0.057 $\pm$ 0.035 & 0.256 $\pm$ 0.010 &\\
9812850   & 1.421 $\pm$ 0.027 & 1.837 $\pm$ 0.012 & 2.352 $\pm$ 0.057 & 0.230 $\pm$ 0.050 & 0.301 $\pm$ 0.015 &\\
9955598   & 0.880 $\pm$ 0.011 & 0.879 $\pm$ 0.004 & 7.048 $\pm$ 0.165 & 0.077 $\pm$ 0.037 & 0.287 $\pm$ 0.011 &\\
9965715   & 1.234 $\pm$ 0.011 & 1.331 $\pm$ 0.005 & 2.644 $\pm$ 0.090 & -0.135 $\pm$ 0.031 & 0.243 $\pm$ 0.003 &\\
10068307  & 1.478 $\pm$ 0.008 & 2.113 $\pm$ 0.005 & 3.232 $\pm$ 0.065 & 0.168 $\pm$ 0.031 & 0.261 $\pm$ 0.006 &\\
10079226  & 1.174 $\pm$ 0.030 & 1.165 $\pm$ 0.010 & 2.458 $\pm$ 0.257 & 0.200 $\pm$ 0.057 & 0.269 $\pm$ 0.018 &\\
10162436  & 1.458 $\pm$ 0.016 & 2.055 $\pm$ 0.008 & 2.732 $\pm$ 0.046 & 0.228 $\pm$ 0.035 & 0.293 $\pm$ 0.010 &\\
10454113  & 1.326 $\pm$ 0.013 & 1.301 $\pm$ 0.004 & 1.426 $\pm$ 0.085 & 0.284 $\pm$ 0.044 & 0.260 $\pm$ 0.009 &\\
10516096  & 1.110 $\pm$ 0.012 & 1.422 $\pm$ 0.006 & 6.389 $\pm$ 0.223 & 0.102 $\pm$ 0.047 & 0.284 $\pm$ 0.011 &\\
10644253  & 1.225 $\pm$ 0.012 & 1.140 $\pm$ 0.004 & 0.971 $\pm$ 0.094 & 0.186 $\pm$ 0.042 & 0.250 $\pm$ 0.007 &\\
10730618  & 1.403 $\pm$ 0.031 & 1.798 $\pm$ 0.015 & 2.216 $\pm$ 0.064 & 0.267 $\pm$ 0.074 & 0.315 $\pm$ 0.017 &\\
10963065  & 1.135 $\pm$ 0.005 & 1.249 $\pm$ 0.002 & 4.307 $\pm$ 0.089 & -0.103 $\pm$ 0.033 & 0.248 $\pm$ 0.006 &\\
11081729  & 1.481 $\pm$ 0.020 & 1.484 $\pm$ 0.008 & 1.027 $\pm$ 0.100 & 0.216 $\pm$ 0.047 & 0.245 $\pm$ 0.005 &\\
11253226  & 1.496 $\pm$ 0.017 & 1.646 $\pm$ 0.006 & 1.040 $\pm$ 0.039 & 0.336 $\pm$ 0.039 & 0.308 $\pm$ 0.011 &\\
11772920  & 0.791 $\pm$ 0.019 & 0.831 $\pm$ 0.007 & 8.655 $\pm$ 0.353 & -0.158 $\pm$ 0.079 & 0.306 $\pm$ 0.018 &\\
12009504  & 1.279 $\pm$ 0.009 & 1.442 $\pm$ 0.004 & 3.494 $\pm$ 0.091 & 0.017 $\pm$ 0.032 & 0.242 $\pm$ 0.002 &\\
12069127  & 1.600 $\pm$ 0.035 & 2.300 $\pm$ 0.018 & 1.868 $\pm$ 0.064 & 0.165 $\pm$ 0.074 & 0.286 $\pm$ 0.021 &\\
12069424  & 1.057 $\pm$ 0.004 & 1.218 $\pm$ 0.001 & 7.201 $\pm$ 0.073 & 0.107 $\pm$ 0.021 & 0.272 $\pm$ 0.004 &\\
12069449  & 1.012 $\pm$ 0.004 & 1.105 $\pm$ 0.002 & 7.108 $\pm$ 0.061 & 0.128 $\pm$ 0.023 & 0.279 $\pm$ 0.005 &\\
12258514  & 1.275 $\pm$ 0.005 & 1.610 $\pm$ 0.002 & 4.529 $\pm$ 0.068 & 0.026 $\pm$ 0.018 & 0.247 $\pm$ 0.003 &\\
12317678  & 1.533 $\pm$ 0.013 & 1.898 $\pm$ 0.005 & 1.965 $\pm$ 0.038 & 0.041 $\pm$ 0.026 & 0.248 $\pm$ 0.007 &\\
\hline
3425851   & 1.348 $\pm$ 0.044 & 1.411 $\pm$ 0.017 & 1.644 $\pm$ 0.155 & 0.119 $\pm$ 0.075 & 0.262 $\pm$ 0.018 \\
3544595   & 0.920 $\pm$ 0.011 & 0.922 $\pm$ 0.004 & 6.679 $\pm$ 0.211 & -0.127 $\pm$ 0.046 & 0.253 $\pm$ 0.010 \\
4141376   & 1.018 $\pm$ 0.036 & 1.037 $\pm$ 0.013 & 3.519 $\pm$ 0.577 & -0.210 $\pm$ 0.073 & 0.264 $\pm$ 0.021 \\
4349452   & 1.265 $\pm$ 0.039 & 1.333 $\pm$ 0.015 & 2.121 $\pm$ 0.155 & 0.100 $\pm$ 0.069 & 0.270 $\pm$ 0.021 \\
4914423   & 1.180 $\pm$ 0.020 & 1.482 $\pm$ 0.009 & 6.701 $\pm$ 0.461 & 0.108 $\pm$ 0.055 & 0.249 $\pm$ 0.009 \\
5094751   & 1.125 $\pm$ 0.028 & 1.363 $\pm$ 0.012 & 5.875 $\pm$ 0.341 & 0.066 $\pm$ 0.060 & 0.271 $\pm$ 0.018 \\
5866724   & 1.345 $\pm$ 0.028 & 1.449 $\pm$ 0.011 & 2.547 $\pm$ 0.129 & 0.191 $\pm$ 0.042 & 0.256 $\pm$ 0.014 \\
6196457   & 1.333 $\pm$ 0.029 & 1.769 $\pm$ 0.014 & 4.411 $\pm$ 0.270 & 0.180 $\pm$ 0.064 & 0.262 $\pm$ 0.017 \\
6278762   & 0.756 $\pm$ 0.010 & 0.753 $\pm$ 0.003 & 11.258 $\pm$ 0.596 & -0.121 $\pm$ 0.045 & 0.265 $\pm$ 0.014 \\
6521045   & 1.148 $\pm$ 0.010 & 1.524 $\pm$ 0.005 & 6.266 $\pm$ 0.164 & -0.075 $\pm$ 0.037 & 0.250 $\pm$ 0.007 \\
7670943   & 1.256 $\pm$ 0.047 & 1.422 $\pm$ 0.020 & 2.479 $\pm$ 0.129 & -0.014 $\pm$ 0.071 & 0.275 $\pm$ 0.023 \\
8077137   & 1.264 $\pm$ 0.026 & 1.690 $\pm$ 0.013 & 3.788 $\pm$ 0.173 & -0.124 $\pm$ 0.058 & 0.260 $\pm$ 0.015 \\
8292840   & 1.231 $\pm$ 0.022 & 1.369 $\pm$ 0.010 & 2.787 $\pm$ 0.114 & -0.142 $\pm$ 0.051 & 0.250 $\pm$ 0.009 \\
8349582   & 1.015 $\pm$ 0.012 & 1.389 $\pm$ 0.006 & 8.092 $\pm$ 0.391 & 0.202 $\pm$ 0.033 & 0.321 $\pm$ 0.007 \\
8478994   & 0.835 $\pm$ 0.007 & 0.780 $\pm$ 0.002 & 4.723 $\pm$ 0.477 & -0.195 $\pm$ 0.041 & 0.246 $\pm$ 0.005 \\
8494142   & 1.153 $\pm$ 0.013 & 1.770 $\pm$ 0.009 & 4.351 $\pm$ 0.158 & 0.158 $\pm$ 0.052 & 0.334 $\pm$ 0.006 \\
8554498   & 1.301 $\pm$ 0.009 & 1.846 $\pm$ 0.005 & 5.817 $\pm$ 0.113 & 0.233 $\pm$ 0.025 & 0.249 $\pm$ 0.006 \\
8866102   & 1.352 $\pm$ 0.022 & 1.399 $\pm$ 0.008 & 1.742 $\pm$ 0.072 & 0.131 $\pm$ 0.038 & 0.253 $\pm$ 0.009 \\
9592705   & 1.499 $\pm$ 0.031 & 2.116 $\pm$ 0.015 & 2.140 $\pm$ 0.046 & 0.344 $\pm$ 0.035 & 0.323 $\pm$ 0.013 \\
10514430  & 1.049 $\pm$ 0.009 & 1.578 $\pm$ 0.005 & 8.167 $\pm$ 0.125 & -0.190 $\pm$ 0.038 & 0.255 $\pm$ 0.005 \\
10586004  & 1.196 $\pm$ 0.026 & 1.658 $\pm$ 0.011 & 7.048 $\pm$ 0.453 & 0.192 $\pm$ 0.061 & 0.259 $\pm$ 0.007 \\
10666592  & 1.569 $\pm$ 0.033 & 2.005 $\pm$ 0.015 & 1.885 $\pm$ 0.050 & 0.182 $\pm$ 0.043 & 0.270 $\pm$ 0.016 \\
11133306  & 1.173 $\pm$ 0.045 & 1.226 $\pm$ 0.016 & 3.499 $\pm$ 0.675 & 0.101 $\pm$ 0.076 & 0.260 $\pm$ 0.017 \\
11295426  & 0.942 $\pm$ 0.011 & 1.141 $\pm$ 0.004 & 6.370 $\pm$ 0.439 & 0.044 $\pm$ 0.034 & 0.335 $\pm$ 0.005 \\
11401755  & 1.125 $\pm$ 0.037 & 1.662 $\pm$ 0.019 & 7.010 $\pm$ 0.351 & -0.094 $\pm$ 0.072 & 0.253 $\pm$ 0.007 \\
11807274  & 1.296 $\pm$ 0.025 & 1.601 $\pm$ 0.011 & 3.298 $\pm$ 0.101 & -0.043 $\pm$ 0.041 & 0.258 $\pm$ 0.014 \\
11853905  & 1.193 $\pm$ 0.016 & 1.583 $\pm$ 0.007 & 6.727 $\pm$ 0.403 & 0.083 $\pm$ 0.040 & 0.247 $\pm$ 0.005 \\
11904151  & 0.911 $\pm$ 0.016 & 1.062 $\pm$ 0.006 & 10.067 $\pm$ 0.367 & -0.022 $\pm$ 0.057 & 0.285 $\pm$ 0.016 \\
\end{longtable}
}

\twocolumn

\nopagebreak
\FloatBarrier
\section{Fundamental properties using 3:3 frequencies weight}

% example for Table A.3:
\onecolumn
% example for Table A.3:
\longtab{
\begin{longtable}{cccccccccccc}
\caption{Inferred fundamental properties of the stars using 3:3 frequencies weight.}
\label{tab:abs_res_33}
\\\hline
KIC       & Mass (M$_\odot$) & Radius (R$_\odot$) & Age (Gyr) &  $\MH_i$ &    $Y_i$ \\
\hline
\endfirsthead
\caption{continued}\\
\hline
KIC       & Mass (M$_\odot$) & Radius (R$_\odot$) & Age (Gyr) &  $\MH_i$ &    $Y_i$ \\
\hline
\endhead
\hline
\endfoot
\hline
\endlastfoot
1435467   & 1.445 $\pm$ 0.024 & 1.740 $\pm$ 0.011 & 2.336 $\pm$ 0.142 & 0.166 $\pm$ 0.063 & 0.267 $\pm$ 0.017 &\\
2837475   & 1.521 $\pm$ 0.025 & 1.672 $\pm$ 0.011 & 1.289 $\pm$ 0.121 & 0.196 $\pm$ 0.073 & 0.269 $\pm$ 0.019 &\\
3427720   & 1.170 $\pm$ 0.014 & 1.139 $\pm$ 0.005 & 2.188 $\pm$ 0.216 & 0.064 $\pm$ 0.047 & 0.249 $\pm$ 0.008 &\\
3456181   & 1.568 $\pm$ 0.030 & 2.184 $\pm$ 0.016 & 2.145 $\pm$ 0.085 & 0.039 $\pm$ 0.060 & 0.255 $\pm$ 0.013 &\\
3632418   & 1.472 $\pm$ 0.013 & 1.945 $\pm$ 0.006 & 2.927 $\pm$ 0.121 & 0.119 $\pm$ 0.039 & 0.253 $\pm$ 0.010 &\\
3656476   & 1.035 $\pm$ 0.009 & 1.296 $\pm$ 0.004 & 8.380 $\pm$ 0.323 & 0.261 $\pm$ 0.030 & 0.306 $\pm$ 0.009 &\\
3735871   & 1.189 $\pm$ 0.021 & 1.127 $\pm$ 0.007 & 1.399 $\pm$ 0.348 & 0.086 $\pm$ 0.053 & 0.249 $\pm$ 0.008 &\\
4914923   & 1.135 $\pm$ 0.012 & 1.385 $\pm$ 0.005 & 6.626 $\pm$ 0.334 & 0.107 $\pm$ 0.052 & 0.262 $\pm$ 0.012 &\\
5184732   & 1.236 $\pm$ 0.009 & 1.352 $\pm$ 0.003 & 4.319 $\pm$ 0.215 & 0.393 $\pm$ 0.021 & 0.282 $\pm$ 0.006 &\\
5773345   & 1.545 $\pm$ 0.028 & 2.043 $\pm$ 0.013 & 2.204 $\pm$ 0.116 & 0.327 $\pm$ 0.056 & 0.291 $\pm$ 0.017 &\\
5950854   & 0.986 $\pm$ 0.017 & 1.245 $\pm$ 0.008 & 9.040 $\pm$ 0.602 & -0.146 $\pm$ 0.060 & 0.261 $\pm$ 0.016 &\\
6106415   & 1.158 $\pm$ 0.007 & 1.250 $\pm$ 0.002 & 4.451 $\pm$ 0.211 & 0.024 $\pm$ 0.026 & 0.247 $\pm$ 0.005 &\\
6116048   & 1.109 $\pm$ 0.009 & 1.261 $\pm$ 0.003 & 5.911 $\pm$ 0.299 & -0.033 $\pm$ 0.033 & 0.248 $\pm$ 0.006 &\\
6225718   & 1.303 $\pm$ 0.008 & 1.289 $\pm$ 0.002 & 1.950 $\pm$ 0.178 & 0.198 $\pm$ 0.025 & 0.244 $\pm$ 0.003 &\\
6508366   & 1.600 $\pm$ 0.021 & 2.219 $\pm$ 0.012 & 2.090 $\pm$ 0.089 & 0.131 $\pm$ 0.060 & 0.262 $\pm$ 0.015 &\\
6603624   & 0.993 $\pm$ 0.007 & 1.142 $\pm$ 0.003 & 8.761 $\pm$ 0.188 & 0.211 $\pm$ 0.026 & 0.296 $\pm$ 0.007 &\\
6679371   & 1.684 $\pm$ 0.024 & 2.281 $\pm$ 0.013 & 1.724 $\pm$ 0.079 & 0.192 $\pm$ 0.063 & 0.265 $\pm$ 0.017 &\\
6933899   & 1.190 $\pm$ 0.005 & 1.617 $\pm$ 0.002 & 6.553 $\pm$ 0.107 & 0.053 $\pm$ 0.012 & 0.249 $\pm$ 0.003 &\\
7103006   & 1.619 $\pm$ 0.023 & 2.025 $\pm$ 0.010 & 2.078 $\pm$ 0.099 & 0.278 $\pm$ 0.047 & 0.257 $\pm$ 0.012 &\\
7206837   & 1.469 $\pm$ 0.025 & 1.622 $\pm$ 0.010 & 2.034 $\pm$ 0.167 & 0.269 $\pm$ 0.057 & 0.258 $\pm$ 0.014 &\\
7296438   & 1.148 $\pm$ 0.016 & 1.390 $\pm$ 0.007 & 6.861 $\pm$ 0.373 & 0.208 $\pm$ 0.049 & 0.267 $\pm$ 0.013 &\\
7510397   & 1.415 $\pm$ 0.013 & 1.886 $\pm$ 0.006 & 3.050 $\pm$ 0.104 & 0.001 $\pm$ 0.042 & 0.250 $\pm$ 0.008 &\\
7680114   & 1.108 $\pm$ 0.012 & 1.411 $\pm$ 0.005 & 7.229 $\pm$ 0.351 & 0.085 $\pm$ 0.056 & 0.267 $\pm$ 0.013 &\\
7771282   & 1.364 $\pm$ 0.047 & 1.678 $\pm$ 0.020 & 2.945 $\pm$ 0.249 & 0.091 $\pm$ 0.070 & 0.266 $\pm$ 0.020 &\\
7871531   & 0.822 $\pm$ 0.015 & 0.866 $\pm$ 0.005 & 9.312 $\pm$ 0.517 & -0.143 $\pm$ 0.043 & 0.288 $\pm$ 0.014 &\\
7940546   & 1.492 $\pm$ 0.013 & 2.001 $\pm$ 0.006 & 2.836 $\pm$ 0.115 & 0.105 $\pm$ 0.034 & 0.251 $\pm$ 0.009 &\\
8006161   & 0.920 $\pm$ 0.006 & 0.907 $\pm$ 0.002 & 5.125 $\pm$ 0.232 & 0.300 $\pm$ 0.038 & 0.318 $\pm$ 0.009 &\\
8150065   & 1.256 $\pm$ 0.045 & 1.422 $\pm$ 0.018 & 3.254 $\pm$ 0.406 & 0.097 $\pm$ 0.084 & 0.269 $\pm$ 0.022 &\\
8179536   & 1.373 $\pm$ 0.021 & 1.400 $\pm$ 0.008 & 1.579 $\pm$ 0.201 & 0.170 $\pm$ 0.054 & 0.250 $\pm$ 0.009 &\\
8228742   & 1.400 $\pm$ 0.011 & 1.887 $\pm$ 0.005 & 3.542 $\pm$ 0.176 & 0.053 $\pm$ 0.040 & 0.248 $\pm$ 0.007 &\\
8379927   & 1.238 $\pm$ 0.009 & 1.163 $\pm$ 0.003 & 1.438 $\pm$ 0.219 & 0.230 $\pm$ 0.032 & 0.247 $\pm$ 0.006 &\\
8394589   & 1.165 $\pm$ 0.012 & 1.212 $\pm$ 0.004 & 3.415 $\pm$ 0.290 & -0.019 $\pm$ 0.041 & 0.246 $\pm$ 0.005 &\\
8424992   & 0.890 $\pm$ 0.021 & 1.035 $\pm$ 0.008 & 9.490 $\pm$ 0.569 & -0.059 $\pm$ 0.070 & 0.296 $\pm$ 0.021 &\\
8938364   & 1.014 $\pm$ 0.006 & 1.361 $\pm$ 0.003 & 9.292 $\pm$ 0.151 & -0.204 $\pm$ 0.025 & 0.244 $\pm$ 0.004 &\\
9025370   & 0.970 $\pm$ 0.018 & 1.000 $\pm$ 0.006 & 4.665 $\pm$ 0.337 & 0.105 $\pm$ 0.088 & 0.304 $\pm$ 0.020 &\\
9098294   & 1.006 $\pm$ 0.011 & 1.156 $\pm$ 0.004 & 7.941 $\pm$ 0.436 & -0.100 $\pm$ 0.049 & 0.253 $\pm$ 0.011 &\\
9139151   & 1.221 $\pm$ 0.014 & 1.172 $\pm$ 0.005 & 1.516 $\pm$ 0.227 & 0.100 $\pm$ 0.044 & 0.248 $\pm$ 0.007 &\\
9139163   & 1.534 $\pm$ 0.013 & 1.617 $\pm$ 0.005 & 1.547 $\pm$ 0.092 & 0.333 $\pm$ 0.040 & 0.250 $\pm$ 0.008 &\\
9206432   & 1.590 $\pm$ 0.024 & 1.577 $\pm$ 0.009 & 0.705 $\pm$ 0.100 & 0.353 $\pm$ 0.048 & 0.251 $\pm$ 0.010 &\\
9353712   & 1.495 $\pm$ 0.040 & 2.169 $\pm$ 0.021 & 2.234 $\pm$ 0.100 & 0.093 $\pm$ 0.055 & 0.287 $\pm$ 0.021 &\\
9410862   & 1.023 $\pm$ 0.021 & 1.175 $\pm$ 0.008 & 6.291 $\pm$ 0.550 & -0.204 $\pm$ 0.062 & 0.256 $\pm$ 0.013 &\\
9414417   & 1.502 $\pm$ 0.019 & 1.967 $\pm$ 0.009 & 2.571 $\pm$ 0.103 & 0.092 $\pm$ 0.048 & 0.251 $\pm$ 0.010 &\\
9812850   & 1.451 $\pm$ 0.027 & 1.848 $\pm$ 0.012 & 2.507 $\pm$ 0.147 & 0.098 $\pm$ 0.064 & 0.264 $\pm$ 0.017 &\\
9955598   & 0.870 $\pm$ 0.015 & 0.876 $\pm$ 0.005 & 6.949 $\pm$ 0.404 & 0.043 $\pm$ 0.066 & 0.292 $\pm$ 0.018 &\\
9965715   & 1.238 $\pm$ 0.013 & 1.333 $\pm$ 0.005 & 2.640 $\pm$ 0.182 & -0.122 $\pm$ 0.029 & 0.242 $\pm$ 0.002 &\\
10068307  & 1.462 $\pm$ 0.011 & 2.100 $\pm$ 0.006 & 3.041 $\pm$ 0.086 & 0.046 $\pm$ 0.042 & 0.255 $\pm$ 0.009 &\\
10079226  & 1.172 $\pm$ 0.040 & 1.164 $\pm$ 0.013 & 2.563 $\pm$ 0.578 & 0.195 $\pm$ 0.068 & 0.268 $\pm$ 0.020 &\\
10162436  & 1.509 $\pm$ 0.015 & 2.076 $\pm$ 0.008 & 2.789 $\pm$ 0.114 & 0.104 $\pm$ 0.042 & 0.252 $\pm$ 0.010 &\\
10454113  & 1.338 $\pm$ 0.013 & 1.303 $\pm$ 0.004 & 1.335 $\pm$ 0.206 & 0.221 $\pm$ 0.041 & 0.247 $\pm$ 0.007 &\\
10516096  & 1.139 $\pm$ 0.010 & 1.433 $\pm$ 0.004 & 6.322 $\pm$ 0.308 & -0.007 $\pm$ 0.048 & 0.255 $\pm$ 0.010 &\\
10644253  & 1.234 $\pm$ 0.017 & 1.142 $\pm$ 0.006 & 0.858 $\pm$ 0.257 & 0.201 $\pm$ 0.049 & 0.249 $\pm$ 0.008 &\\
10730618  & 1.415 $\pm$ 0.040 & 1.803 $\pm$ 0.019 & 2.384 $\pm$ 0.201 & 0.175 $\pm$ 0.110 & 0.290 $\pm$ 0.026 &\\
10963065  & 1.156 $\pm$ 0.009 & 1.258 $\pm$ 0.003 & 4.223 $\pm$ 0.267 & -0.029 $\pm$ 0.040 & 0.248 $\pm$ 0.007 &\\
11081729  & 1.490 $\pm$ 0.023 & 1.486 $\pm$ 0.009 & 0.897 $\pm$ 0.154 & 0.250 $\pm$ 0.056 & 0.251 $\pm$ 0.009 &\\
11253226  & 1.524 $\pm$ 0.025 & 1.657 $\pm$ 0.010 & 1.273 $\pm$ 0.108 & 0.173 $\pm$ 0.065 & 0.262 $\pm$ 0.016 &\\
11772920  & 0.803 $\pm$ 0.019 & 0.836 $\pm$ 0.007 & 9.310 $\pm$ 0.687 & 0.037 $\pm$ 0.083 & 0.312 $\pm$ 0.020 &\\
12009504  & 1.308 $\pm$ 0.012 & 1.455 $\pm$ 0.005 & 3.409 $\pm$ 0.220 & 0.118 $\pm$ 0.033 & 0.245 $\pm$ 0.005 &\\
12069127  & 1.603 $\pm$ 0.046 & 2.307 $\pm$ 0.024 & 1.941 $\pm$ 0.111 & 0.182 $\pm$ 0.078 & 0.284 $\pm$ 0.024 &\\
12069424  & 1.044 $\pm$ 0.007 & 1.213 $\pm$ 0.003 & 7.092 $\pm$ 0.257 & 0.103 $\pm$ 0.039 & 0.280 $\pm$ 0.010 &\\
12069449  & 0.998 $\pm$ 0.007 & 1.099 $\pm$ 0.003 & 7.002 $\pm$ 0.208 & 0.098 $\pm$ 0.038 & 0.284 $\pm$ 0.009 &\\
12258514  & 1.303 $\pm$ 0.009 & 1.625 $\pm$ 0.003 & 4.714 $\pm$ 0.211 & 0.122 $\pm$ 0.026 & 0.245 $\pm$ 0.004 &\\
12317678  & 1.477 $\pm$ 0.018 & 1.875 $\pm$ 0.009 & 1.995 $\pm$ 0.117 & 0.009 $\pm$ 0.077 & 0.264 $\pm$ 0.018 &\\
\hline
3425851   & 1.331 $\pm$ 0.051 & 1.408 $\pm$ 0.018 & 1.897 $\pm$ 0.358 & 0.114 $\pm$ 0.079 & 0.262 $\pm$ 0.019 \\
3544595   & 0.904 $\pm$ 0.022 & 0.916 $\pm$ 0.008 & 6.637 $\pm$ 0.548 & -0.138 $\pm$ 0.064 & 0.265 $\pm$ 0.018 \\
4141376   & 1.023 $\pm$ 0.043 & 1.039 $\pm$ 0.015 & 3.449 $\pm$ 0.929 & -0.194 $\pm$ 0.084 & 0.265 $\pm$ 0.021 \\
4349452   & 1.248 $\pm$ 0.050 & 1.328 $\pm$ 0.018 & 2.345 $\pm$ 0.383 & 0.077 $\pm$ 0.074 & 0.270 $\pm$ 0.023 \\
4914423   & 1.168 $\pm$ 0.035 & 1.477 $\pm$ 0.015 & 6.585 $\pm$ 0.719 & 0.111 $\pm$ 0.062 & 0.258 $\pm$ 0.017 \\
5094751   & 1.101 $\pm$ 0.044 & 1.352 $\pm$ 0.018 & 5.962 $\pm$ 0.651 & 0.015 $\pm$ 0.068 & 0.276 $\pm$ 0.025 \\
5866724   & 1.293 $\pm$ 0.052 & 1.431 $\pm$ 0.019 & 2.860 $\pm$ 0.346 & 0.172 $\pm$ 0.061 & 0.270 $\pm$ 0.023 \\
6196457   & 1.322 $\pm$ 0.048 & 1.762 $\pm$ 0.023 & 4.204 $\pm$ 0.408 & 0.215 $\pm$ 0.071 & 0.277 $\pm$ 0.026 \\
6278762   & 0.752 $\pm$ 0.017 & 0.751 $\pm$ 0.006 & 11.794 $\pm$ 1.457 & -0.220 $\pm$ 0.059 & 0.254 $\pm$ 0.013 \\
6521045   & 1.098 $\pm$ 0.022 & 1.504 $\pm$ 0.011 & 6.682 $\pm$ 0.387 & 0.070 $\pm$ 0.050 & 0.287 $\pm$ 0.015 \\
7670943   & 1.286 $\pm$ 0.058 & 1.435 $\pm$ 0.023 & 2.348 $\pm$ 0.325 & 0.084 $\pm$ 0.084 & 0.278 $\pm$ 0.026 \\
8077137   & 1.262 $\pm$ 0.040 & 1.692 $\pm$ 0.019 & 3.857 $\pm$ 0.332 & -0.055 $\pm$ 0.056 & 0.269 $\pm$ 0.022 \\
8292840   & 1.222 $\pm$ 0.030 & 1.368 $\pm$ 0.012 & 3.068 $\pm$ 0.346 & -0.118 $\pm$ 0.059 & 0.252 $\pm$ 0.011 \\
8349582   & 1.040 $\pm$ 0.025 & 1.402 $\pm$ 0.012 & 8.525 $\pm$ 0.654 & 0.260 $\pm$ 0.052 & 0.312 $\pm$ 0.015 \\
8478994   & 0.854 $\pm$ 0.016 & 0.786 $\pm$ 0.005 & 3.505 $\pm$ 1.014 & -0.173 $\pm$ 0.044 & 0.245 $\pm$ 0.005 \\
8494142   & 1.181 $\pm$ 0.072 & 1.782 $\pm$ 0.034 & 4.224 $\pm$ 0.438 & 0.135 $\pm$ 0.062 & 0.324 $\pm$ 0.020 \\
8554498   & 1.249 $\pm$ 0.039 & 1.818 $\pm$ 0.018 & 5.298 $\pm$ 0.413 & 0.196 $\pm$ 0.041 & 0.276 $\pm$ 0.017 \\
8866102   & 1.322 $\pm$ 0.037 & 1.390 $\pm$ 0.014 & 1.892 $\pm$ 0.234 & 0.139 $\pm$ 0.064 & 0.265 $\pm$ 0.019 \\
9592705   & 1.548 $\pm$ 0.057 & 2.140 $\pm$ 0.027 & 2.251 $\pm$ 0.133 & 0.291 $\pm$ 0.061 & 0.293 $\pm$ 0.026 \\
10514430  & 1.046 $\pm$ 0.013 & 1.578 $\pm$ 0.007 & 8.412 $\pm$ 0.317 & -0.164 $\pm$ 0.049 & 0.256 $\pm$ 0.007 \\
10586004  & 1.266 $\pm$ 0.030 & 1.689 $\pm$ 0.015 & 6.389 $\pm$ 0.924 & 0.281 $\pm$ 0.056 & 0.257 $\pm$ 0.017 \\
10666592  & 1.582 $\pm$ 0.063 & 2.015 $\pm$ 0.028 & 1.865 $\pm$ 0.098 & 0.295 $\pm$ 0.057 & 0.284 $\pm$ 0.027 \\
11133306  & 1.117 $\pm$ 0.054 & 1.207 $\pm$ 0.020 & 4.552 $\pm$ 0.988 & 0.048 $\pm$ 0.081 & 0.265 $\pm$ 0.020 \\
11295426  & 0.906 $\pm$ 0.033 & 1.130 $\pm$ 0.020 & 8.080 $\pm$ 0.829 & 0.035 $\pm$ 0.046 & 0.332 $\pm$ 0.011 \\
11401755  & 1.072 $\pm$ 0.017 & 1.634 $\pm$ 0.009 & 7.229 $\pm$ 0.379 & -0.219 $\pm$ 0.033 & 0.257 $\pm$ 0.011 \\
11807274  & 1.276 $\pm$ 0.053 & 1.594 $\pm$ 0.023 & 3.310 $\pm$ 0.254 & 0.027 $\pm$ 0.061 & 0.278 $\pm$ 0.026 \\
11853905  & 1.184 $\pm$ 0.030 & 1.580 $\pm$ 0.013 & 6.924 $\pm$ 0.655 & 0.096 $\pm$ 0.054 & 0.250 $\pm$ 0.010 \\
11904151  & 0.912 $\pm$ 0.028 & 1.063 $\pm$ 0.011 & 10.276 $\pm$ 0.751 & -0.048 $\pm$ 0.070 & 0.280 $\pm$ 0.023 \\
\end{longtable}
}% End longtab 

\twocolumn
\end{appendix}
%\newpage\phantom{---}
%\newpage

\end{document}